\pdfoutput=1
\documentclass[12pt,a4paper]{article}

\usepackage{ifthen} 
\newboolean{pdflatex}
\setboolean{pdflatex}{true} 

\newboolean{articletitles}
\setboolean{articletitles}{true} 

\newboolean{uprightparticles}
\setboolean{uprightparticles}{false} 

\usepackage{hyperref}    
\usepackage{cite} 
\usepackage{mciteplus}

\usepackage{overpic}

\usepackage{amsmath}

\usepackage{lineno}  
\usepackage{xspace} 

\usepackage{graphicx}  
\usepackage{color}
\usepackage{colortbl}
\graphicspath{{./figs/}} 

\usepackage{amsmath} 
\usepackage{amssymb}
\usepackage{amsfonts}
\usepackage{upgreek} 

\newcommand*\patchAmsMathEnvironmentForLineno[1]{%
\expandafter\let\csname old#1\expandafter\endcsname\csname #1\endcsname
\expandafter\let\csname oldend#1\expandafter\endcsname\csname
end#1\endcsname
 \renewenvironment{#1}%
   {\linenomath\csname old#1\endcsname}%
   {\csname oldend#1\endcsname\endlinenomath}%
}
\newcommand*\patchBothAmsMathEnvironmentsForLineno[1]{%
  \patchAmsMathEnvironmentForLineno{#1}%
  \patchAmsMathEnvironmentForLineno{#1*}%
}
\AtBeginDocument{%
\patchBothAmsMathEnvironmentsForLineno{equation}%
\patchBothAmsMathEnvironmentsForLineno{align}%
\patchBothAmsMathEnvironmentsForLineno{flalign}%
\patchBothAmsMathEnvironmentsForLineno{alignat}%
\patchBothAmsMathEnvironmentsForLineno{gather}%
\patchBothAmsMathEnvironmentsForLineno{multline}%
}

\usepackage{hyperref}    
\usepackage[all]{hypcap} 

\def\lhcb {\mbox{LHCb}\xspace}
\def\ux85 {\mbox{UX85}\xspace}

\ifthenelse{\boolean{uprightparticles}}%
{

 \def\Pmu         {\ensuremath{\upmu}\xspace}

 \def\Ppi         {\ensuremath{\uppi}\xspace}

 \def\Ppsi        {\ensuremath{\uppsi}\xspace}

 \def\PDelta      {\ensuremath{\Delta}\xspace}                 
 \def\PXi      {\ensuremath{\Xi}\xspace}                 
 \def\PLambda      {\ensuremath{\Lambda}\xspace}                 
 \def\PSigma      {\ensuremath{\Sigma}\xspace}                 
 \def\POmega      {\ensuremath{\Omega}\xspace}                 
 \def\PUpsilon      {\ensuremath{\Upsilon}\xspace}                 
 
 \def\PB      {\ensuremath{\mathrm{B}}\xspace}                 
                  
 \def\PD      {\ensuremath{\mathrm{D}}\xspace}

 \def\PJ      {\ensuremath{\mathrm{J}}\xspace}                 
 \def\PK      {\ensuremath{\mathrm{K}}\xspace}

 \def\Pb      {\ensuremath{\mathrm{b}}\xspace}                 
 \def\Pc      {\ensuremath{\mathrm{c}}\xspace}

 \def\Pi      {\ensuremath{\mathrm{i}}\xspace}

 \def\Ps      {\ensuremath{\mathrm{s}}\xspace}

}
{

 \def\Pmu         {\ensuremath{\mu}\xspace}

 \def\Ppi         {\ensuremath{\pi}\xspace}

 \def\Ppsi        {\ensuremath{\psi}\xspace}                 
                  
 \mathchardef\PDelta="7101
 \mathchardef\PXi="7104
 \mathchardef\PLambda="7103
 \mathchardef\PSigma="7106
 \mathchardef\POmega="710A
 \mathchardef\PUpsilon="7107
                  
 \def\PB      {\ensuremath{B}\xspace}                 
                  
 \def\PD      {\ensuremath{D}\xspace}

 \def\PJ      {\ensuremath{J}\xspace}                 
 \def\PK      {\ensuremath{K}\xspace}

 \def\Pb      {\ensuremath{b}\xspace}                 
 \def\Pc      {\ensuremath{c}\xspace}

 \def\Pi      {\ensuremath{i}\xspace}

 \def\Ps      {\ensuremath{s}\xspace}

}

\def\mup        {\ensuremath{\Pmu^+}\xspace}
\def\mun        {\ensuremath{\Pmu^-}\xspace} 

\def\squark    {\ensuremath{\Ps}\xspace}
\def\squarkbar {\ensuremath{\overline \squark}\xspace}

\def\cquark    {\ensuremath{\Pc}\xspace}

\def\bquark    {\ensuremath{\Pb}\xspace}
\def\bquarkbar {\ensuremath{\overline \bquark}\xspace}

\def\pion  {\ensuremath{\Ppi}\xspace}

\def\pip   {\ensuremath{\pion^+}\xspace}
\def\pim   {\ensuremath{\pion^-}\xspace}

\def\kaon  {\ensuremath{\PK}\xspace}
  \def\Kbar  {\kern 0.2em\overline{\kern -0.2em \PK}{}\xspace}

\def\Kz    {\ensuremath{\kaon^0}\xspace}
\def\Kzb   {\ensuremath{\Kbar^0}\xspace}
\def\KzKzb {\ensuremath{\Kz \kern -0.16em \Kzb}\xspace}
\def\Kp    {\ensuremath{\kaon^+}\xspace}
\def\Km    {\ensuremath{\kaon^-}\xspace}

\def\KpKm  {\ensuremath{\Kp \kern -0.16em \Km}\xspace}

\def\Kstarz  {\ensuremath{\kaon^{*0}}\xspace}

\def\Kstar   {\ensuremath{\kaon^*}\xspace}

  \def\Dbar    {\kern 0.2em\overline{\kern -0.2em \PD}{}\xspace}
\def\D       {\ensuremath{\PD}\xspace}

\def\Dz      {\ensuremath{\D^0}\xspace}
\def\Dzb     {\ensuremath{\Dbar^0}\xspace}
\def\DzDzb   {\ensuremath{\Dz {\kern -0.16em \Dzb}}\xspace}
\def\Dp      {\ensuremath{\D^+}\xspace}
\def\Dm      {\ensuremath{\D^-}\xspace}

\def\DpDm    {\ensuremath{\Dp {\kern -0.16em \Dm}}\xspace}

\def\Dsm     {\ensuremath{\D^-_\squark}\xspace}

\def\B       {\ensuremath{\PB}\xspace}
  \def\Bbar    {\kern 0.18em\overline{\kern -0.18em \PB}{}\xspace}
\def\Bb      {\ensuremath{\Bbar}\xspace}

\def\Bu      {\ensuremath{\B^+}\xspace}

\def\Bd      {\ensuremath{\B^0}\xspace}
\def\Bs      {\ensuremath{\B^0_\squark}\xspace}
\def\Bsb     {\ensuremath{\Bbar^0_\squark}\xspace}

\def\jpsi     {\ensuremath{{\PJ\mskip -3mu/\mskip -2mu\Ppsi\mskip 2mu}}\xspace}

  \def\Y#1S{\ensuremath{\PUpsilon{(#1S)}}\xspace}

\def\Lbar {\ensuremath{\kern 0.1em\overline{\kern -0.1em\PLambda}}\xspace}

\newcommand{\decay}[2]{\ensuremath{#1\!\to #2}\xspace}         

\def\to                 {\ensuremath{\rightarrow}\xspace}

\newcommand{\mBs}{\ensuremath{m_{\Bs}}\xspace}

\def\CP                {\ensuremath{C\!P}\xspace}

\newcommand{\dms}{\ensuremath{\Delta m_{\squark}}\xspace}

\newcommand{\DGs}{\ensuremath{\Delta\Gamma_{\squark}}\xspace}

\newcommand{\Gs}{\ensuremath{\Gamma_{\squark}}\xspace}

\newcommand{\GH}{\ensuremath{\Gamma_{\rm H}}\xspace}

\newcommand{\phis}{\ensuremath{\phi_{\squark}}\xspace}
\newcommand{\betas}{\ensuremath{\beta_{\squark}}\xspace}

\newcommand{\mistag}{\ensuremath{\omega}\xspace}

\newcommand{\etag}{{\ensuremath{\varepsilon_{\rm tag}}}\xspace}

\newcommand{\effeff}{\ensuremath{\varepsilon_{\rm eff}}\xspace}

\newcommand{\effD}{{\ensuremath{\etag {\cal D}^2}}\xspace}

\def\BsToJPsiPhi  {\decay{\Bs}{\jpsi\phi}}
\def\BdToJPsiKst  {\decay{\Bd}{\jpsi\Kstarz}}

\def\BuToJPsiK  {\decay{\Bu}{\jpsi\Kp}}

\def\BsToDsPi  {\decay{\Bs}{\Dsm\pip}}

\def\AT#1     {\ensuremath{A_{\mathrm{T}}^{#1}}\xspace}           

\def\C#1      {\ensuremath{\mathcal{C}_{#1}}\xspace}                       
\def\Cp#1     {\ensuremath{\mathcal{C}_{#1}^{'}}\xspace}                    
\def\Ceff#1   {\ensuremath{\mathcal{C}_{#1}^{\mathrm{(eff)}}}\xspace}        
\def\Cpeff#1  {\ensuremath{\mathcal{C}_{#1}^{'\mathrm{(eff)}}}\xspace}       
\def\Ope#1    {\ensuremath{\mathcal{O}_{#1}}\xspace}                       
\def\Opep#1   {\ensuremath{\mathcal{O}_{#1}^{'}}\xspace}                    

\newcommand{\braket}[2]{\ensuremath{\langle #1|#2\rangle}} 

\newcommand{\tev}{\ensuremath{\mathrm{\,Te\kern -0.1em V}}\xspace}
\newcommand{\gev}{\ensuremath{\mathrm{\,Ge\kern -0.1em V}}\xspace}
\newcommand{\mev}{\ensuremath{\mathrm{\,Me\kern -0.1em V}}\xspace}
\newcommand{\kev}{\ensuremath{\mathrm{\,ke\kern -0.1em V}}\xspace}
\newcommand{\ev}{\ensuremath{\mathrm{\,e\kern -0.1em V}}\xspace}
\newcommand{\gevc}{\ensuremath{{\mathrm{\,Ge\kern -0.1em V\!/}c}}\xspace}
\newcommand{\mevc}{\ensuremath{{\mathrm{\,Me\kern -0.1em V\!/}c}}\xspace}
\newcommand{\gevcc}{\ensuremath{{\mathrm{\,Ge\kern -0.1em V\!/}c^2}}\xspace}
\newcommand{\gevgevcccc}{\ensuremath{{\mathrm{\,Ge\kern -0.1em V^2\!/}c^4}}\xspace}
\newcommand{\mevcc}{\ensuremath{{\mathrm{\,Me\kern -0.1em V\!/}c^2}}\xspace}

\def\mum  {\ensuremath{\,\upmu\rm m}\xspace}

\def\invfb   {\ensuremath{\mbox{\,fb}^{-1}}\xspace}

\def\ps   {\ensuremath{{\rm \,ps}}\xspace}

\def\invps{\ensuremath{{\rm \,ps^{-1}}}\xspace}

\def\deriv {\ensuremath{\mathrm{d}}}

\def\gsim{{~\raise.15em\hbox{$>$}\kern-.85em
          \lower.35em\hbox{$\sim$}~}\xspace}
\def\lsim{{~\raise.15em\hbox{$<$}\kern-.85em
          \lower.35em\hbox{$\sim$}~}\xspace}

\def\sPlot{\mbox{\em sPlot}}

\def\pt         {\mbox{$p_{\rm T}$}\xspace}

\def\rad{\ensuremath{\rm \,rad}\xspace}

\def\evtgen     {\mbox{\textsc{EvtGen}}\xspace}
\def\pythia     {\mbox{\textsc{Pythia}}\xspace}

\def\geant      {\mbox{\textsc{Geant4}}\xspace}

\def\photos     {\mbox{\textsc{Photos}}\xspace}

\def\tell1  {TELL1\xspace}
\def\ukl1   {UKL1\xspace}

\usepackage{cite} 
\usepackage{mciteplus}

\begin{document}

\renewcommand{\thefootnote}{\fnsymbol{footnote}}
\setcounter{footnote}{1}


%
%

\def\BtoJpsiKK{B_s^0 \to J/\psi K^+K^-}
\def\BtoJpsipipi{B_s^0 \to J/\psi \pi^+\pi^-}

\def\bl{B_{\text L}}
\def\bh{B_{\text H}}

\def\delone{\delta_1}
\def\deltwo{\delta_2}
\def\delpar{\delta_\parallel}
\def\delperp{\delta_\perp}
\def\delzero{\delta_0}
\def\dels{\delta_{\rm S}}
\def\delsperp{\dels - \delperp}

\def\apar{A_{\|}}
\def\aperp{A_{\perp}}
\def\aparsq{|A_{\|}|^2}
\def\aperpsq{|A_{\perp}|^2}
\def\assq{|A_{s}|^2}
\def\azero{A_{0}(t)}
\def\azerosq{|A_{0}|^2}

\def\aparsqo{|A_{\|}(0)|^2}
\def\aperpsqo{|A_{\perp}(0)|^2}
\def\assqo{|A_{s}(0)|^2}
\def\azerosqo{|A_{0}(0)|^2}

\def\fS{F_{\text{S}}}
\def\maglambda{|\lambda|}

\def\thetatr{\theta_{tr}}
\def\phitr{\phi_{tr}}
\def\psitr{\psi_{tr}}
\def\thetaone{\theta_{\phi}}
\def\thetatwo{\theta_2}

\def\thetamu{\ensuremath{\theta_{\mu}}\xspace}
\def\thetaK{\ensuremath{\theta_{K}}\xspace}
\def\phihel{\ensuremath{\varphi_{h}}\xspace}

\def\alphab{\alpha_{\text{bkg}}}

\def\tagomega{\eta}
\def\wtrue{\omega}
\def\wtruebar{\bar\omega}
\def\effmistag{\sqrt{\varepsilon_{tag}} (1 - 2 w) }
\def\eventres{\sigma_t}
\def\eventresscale{r_t}

\def\LLF{{\cal L}}
\def\fullpdf{{\cal P}}
\def\sigpdf{{\cal S}}
\def\bkgpdf{{\cal B}}
\def\fsig{f_{s}}


\begin{titlepage}
\pagenumbering{roman}

\vspace*{-1.5cm}
\centerline{\large EUROPEAN ORGANIZATION FOR NUCLEAR RESEARCH (CERN)}
\vspace*{1.5cm}
\hspace*{-0.5cm}
\begin{tabular*}{\linewidth}{lc@{\extracolsep{\fill}}r}
\ifthenelse{\boolean{pdflatex}}
{\vspace*{-2.7cm}\mbox{\!\!\!\includegraphics[width=.14\textwidth]{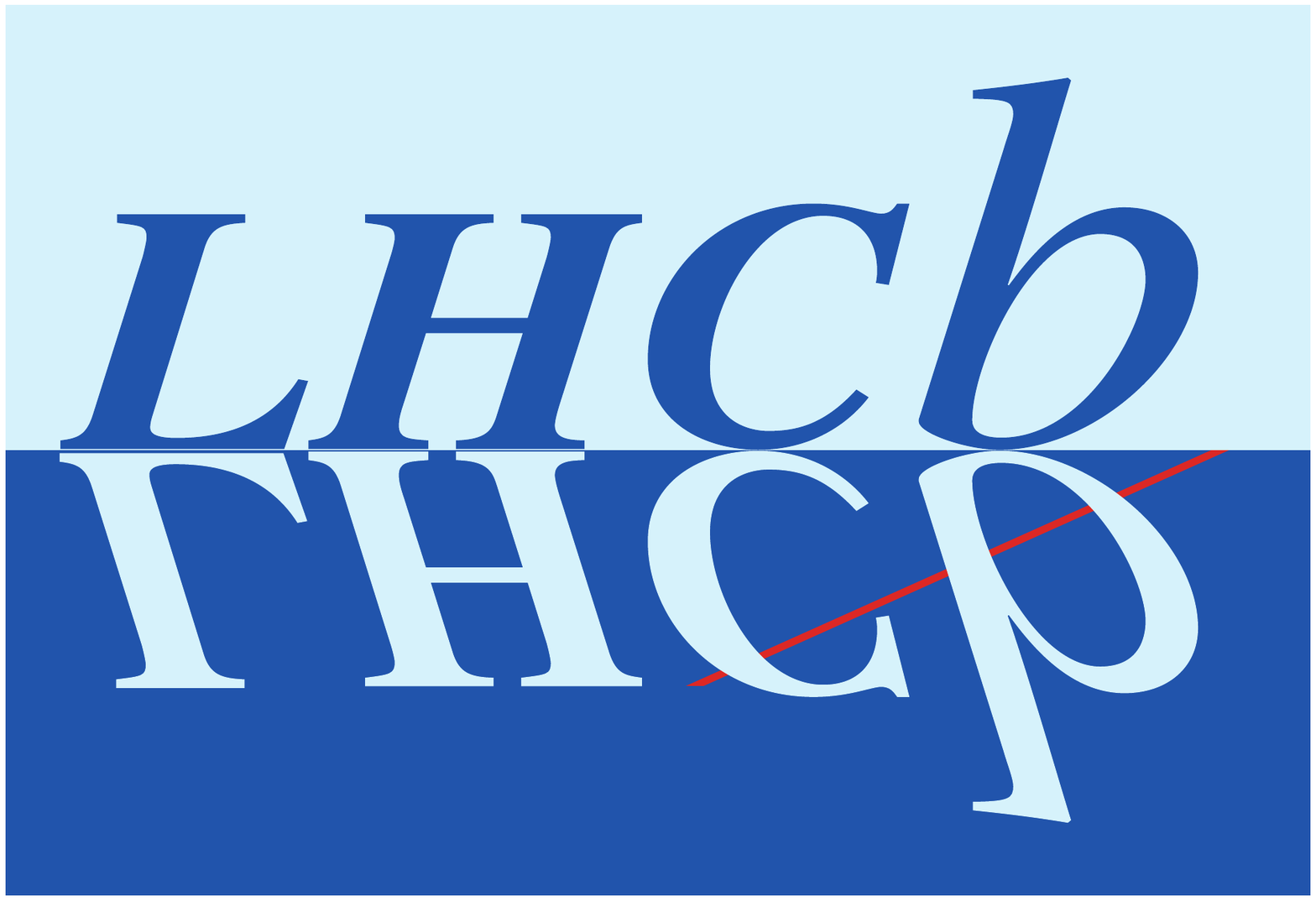}} & &}%
{\vspace*{-1.2cm}\mbox{\!\!\!\includegraphics[width=.12\textwidth]{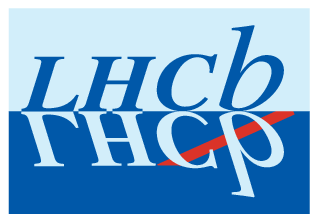}} & &}%
\\
 & & CERN-PH-EP-2013-055 \\  
 & & LHCb-PAPER-2013-002 \\  
 & & May 22, 2013 \\ 
 & & \\
\end{tabular*}


{\bf\boldmath\huge
\begin{center}
Measurement of \CP{} violation and the $\Bs$ meson decay width difference with $\BtoJpsiKK$ and $\BtoJpsipipi$ decays
\end{center}
}

\vspace*{.0cm}

\begin{center}
The LHCb collaboration\footnote{Authors are listed on the following pages.}
\end{center}


\begin{abstract}
  \noindent
  The time-dependent \CP\ asymmetry in  $\Bs\to J/\psi K^{+}K^{-}$ decays is measured
  using $pp$ collision data at $\sqrt{s}=7 \tev$, corresponding to an integrated
  luminosity of $1.0$\invfb, collected with the LHCb detector. 
  The decay time distribution 
  is characterised by the decay widths
  $\Gamma_{\mathrm{L}}$ and $\Gamma_{\mathrm{H}}$ of the light and
  heavy mass eigenstates of the \Bs--\Bsb system and by a
  \CP-violating phase \phis.  In a sample of 27\,617
  $\Bs\to J/\psi K^{+}K^{-}$ decays, where the dominant contribution comes 
  from $\Bs\to J/\psi\phi$ decays, these parameters are measured to be 
  $\phi_s = 0.07  \pm 0.09  \text{(stat)} \pm 0.01 \text{(syst)}\ \text{rad}$,
  $\Gamma_s \equiv (\Gamma_{\mathrm{L}}+\Gamma_{\mathrm{H}})/2 
  = 0.663  \pm  0.005  \text{(stat)} \pm 0.006 \text{(syst)}\ \invps$ and
  $\DGs   \equiv \Gamma_{\mathrm{L}} - \Gamma_{\mathrm{H}}  
  = 0.100   \pm  0.016    \text{(stat)} \pm  0.003  \text{(syst)}\ \invps$,
corresponding to the single most precise determination of \phis, $\Delta\Gamma_{s}$ and $\Gamma_{s}$.
The result of performing a combined analysis with $B_s^{0} \rightarrow J/\psi \pi^{+}\pi^{-}$ decays gives
$\phi_s = 0.01  \pm  0.07   \text{(stat)} \pm 0.01 \text{(syst)}\ \text{rad}$,
$\Gamma_s = 0.661  \pm  0.004  \text{(stat)}\pm 0.006  \text{(syst)}\ \invps$ and
$\DGs  = 0.106   \pm  0.011    \text{(stat)} \pm 0.007 \text{(syst)}\ \invps$.
All measurements are in agreement with the Standard Model predictions.
\end{abstract}

\vspace*{.0cm}

\begin{center}
  Submitted to Phys. Rev. D
\end{center}

\vspace{\fill}

{\footnotesize 
    \centerline{\copyright~CERN on behalf of the \lhcb collaboration, license \href{http://creativecommons.org/licenses/by/3.0/}{CC-BY-3.0}.}}
\vspace*{2mm}

\end{titlepage}


\newpage
\setcounter{page}{2}
\mbox{~}
\newpage

\centerline{\large\bf LHCb collaboration}
\begin{flushleft}
\small
R.~Aaij$^{40}$, 
C.~Abellan~Beteta$^{35,n}$, 
B.~Adeva$^{36}$, 
M.~Adinolfi$^{45}$, 
C.~Adrover$^{6}$, 
A.~Affolder$^{51}$, 
Z.~Ajaltouni$^{5}$, 
J.~Albrecht$^{9}$, 
F.~Alessio$^{37}$, 
M.~Alexander$^{50}$
% ...

{\footnotesize \it
% ...
$ ^{5}$Clermont Universit\'{e}, Universit\'{e} Blaise Pascal, CNRS/IN2P3, LPC, Clermont-Ferrand, France\\
$ ^{6}$CPPM, Aix-Marseille Universit\'{e}, CNRS/IN2P3, Marseille, France\\
% ...
$ ^{9}$Fakult\"{a}t Physik, Technische Universit\"{a}t Dortmund, Dortmund, Germany\\
% ...
$ ^{35}$Universitat de Barcelona, Barcelona, Spain\\
$ ^{36}$Universidad de Santiago de Compostela, Santiago de Compostela, Spain\\
$ ^{37}$European Organization for Nuclear Research (CERN), Geneva, Switzerland\\
% ...
$ ^{40}$Nikhef National Institute for Subatomic Physics, Amsterdam, The Netherlands\\
% ...
$ ^{45}$H.H. Wills Physics Laboratory, University of Bristol, Bristol, United Kingdom\\
% ...
$ ^{50}$School of Physics and Astronomy, University of Glasgow, Glasgow, United Kingdom\\
$ ^{51}$Oliver Lodge Laboratory, University of Liverpool, Liverpool, United Kingdom\\
% ...
$ ^{n}$LIFAELS, La Salle, Universitat Ramon Llull, Barcelona, Spain\\
% ...
$ ^{s}$Scuola Normale Superiore, Pisa, Italy\\
}
\end{flushleft}

\cleardoublepage


\renewcommand{\thefootnote}{\arabic{footnote}}
\setcounter{footnote}{0}



\pagestyle{plain} 
\setcounter{page}{1}
\pagenumbering{arabic}


%

%
%

\section{Introduction \label{sec:introduction}}

\begin{figure}[t]
\begin{center}
\includegraphics*[scale=1, clip=true, trim=30mm 200mm 0mm 40mm]{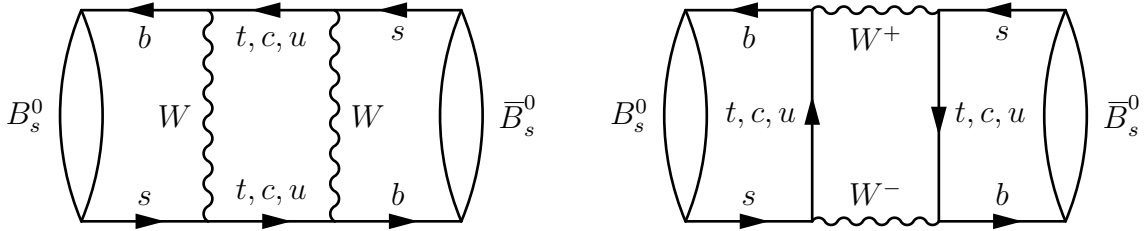}
\end{center}\vspace{-0.4cm}
\caption{\small Feynman diagrams for \Bs--\Bsb\ mixing, within the SM.}
\label{fig:boxOL}
\end{figure}

The interference between \Bs meson decay amplitudes to    \CP eigenstates $\jpsi X$ directly or via mixing 
gives rise to a measurable \CP-violating phase \phis.
In the Standard Model (SM), for $b\to c\overline{c}s$ transitions and ignoring subleading penguin contributions, this phase is predicted to be $-2\betas$, where $\betas=\arg\left(                                 
- V_{ts} V_{tb}^*/ V_{cs} V_{cb}^*\right)$ and $V_{ij}$ are elements of the CKM quark flavour mixing
matrix~\cite{Kobayashi:1973fv,*Cabibbo:1963yz}. The indirect determination via global fits to experimental data gives
\mbox{$2\betas=0.0364\pm0.0016\rad$}~\cite{CKMfitter}. This precise indirect determination within the SM makes the
measurement of \phis\ interesting since new physics (NP) processes could modify 
the phase if new particles were to contribute to the \Bs--\Bsb\ box diagrams~\cite{Buras:2009if,Chiang:2009ev} shown
in Fig.~\ref{fig:boxOL}.

Direct measurements of \phis\ using \BsToJPsiPhi\ and $\BtoJpsipipi$ decays have been reported previously. In the \BsToJPsiPhi\ channel, 
the decay width difference of the light (L) and heavy (H) \Bs\ mass eigenstates,   
$\DGs \equiv \Gamma_{\mathrm L} - \Gamma_{\mathrm H}$, and the average
\Bs-decay width, $\Gs = (\Gamma_{\mathrm L} +\Gamma_{\mathrm H})/2$ are also measured. The measurements of
\phis\ and \DGs\ are shown in Table~\ref{tab:summary}.

This paper extends previous \lhcb measurements in the \BsToJPsiPhi\  
\cite{LHCb:2011aa} and \mbox{$\BtoJpsipipi$}~\cite{LHCb-PAPER-2012-006} channels. 
In the previous analysis of \BsToJPsiPhi\ decays, the invariant mass of the  $\Kp\Km$ system was limited to $\pm 12 \mevcc$ around the $\phi(1020)$ mass~\cite{pdg}, 
which selected predominately resonant P-wave $\phi\to\Kp\Km$ events, although a small S-wave $\Kp\Km$ component was also present. 
In this analysis the $\Kp\Km$ mass range is extended to $\pm 30 \mevcc$ and the notation $\Bs\to\jpsi\Kp\Km$ is used to include explicitly both \mbox{P- and} S-wave decays~\cite{Aaij:2013orb}. 
In both channels additional same-side flavour tagging information is used.
The data were obtained from $pp$ collisions collected by the 
LHCb experiment at a centre-of-mass energy of 7\tev during 2011,
corresponding to an integrated luminosity of $1.0\invfb$.

\begin{table}[b]\footnotesize
\caption{\small Results for $\phis$ and $\Delta\Gamma_{s}$ from different experiments. 
The first uncertainty is statistical and the second is systematic (apart from the D0 result, for which the uncertainties are combined). The CDF confidence level (CL) range quoted is that 
consistent with other experimental measurements of \phis.\label{tab:summary}}
\begin{tabular}{lcccc}
Experiment			&	Dataset [\invfb]		& Ref.					&	$\phis$[\rad]	&	$\Delta\Gamma_{s}$[\invps]	\\
\hline
LHCb (\BsToJPsiPhi)	&	$0.4$			& \cite{LHCb:2011aa}			&	$0.15 \pm 0.18 \pm 0.06$	& $0.123 \pm 0.029 \pm 0.011$\\
LHCb ($\BtoJpsipipi$)	&	$1.0$			& \cite{LHCb-PAPER-2012-006}	&	$-0.019\,^{+0.173+0.004}_{-0.174-0.003}$	&	--\\
LHCb (combined)		&	$0.4$+$1.0$	& \cite{LHCb-PAPER-2012-006}	& 	$0.06 \pm 0.12 \pm 0.06$	& --\\
ATLAS				&	$4.9$			& \cite{atlas:2012fu}				&	$0.22 \pm 0.41 \pm 0.10$	&  $0.053 \pm 0.021 \pm 0.010$\\
CMS					& 	$5.0$				& \cite{CMS:2012}				&	--						&  $0.048 \pm 0.024 \pm 0.003$\\
D0					&	$8.0$			& \cite{Abazov:2011ry}			&	$-0.55\,^{+0.38} _{-0.36}$		&  $0.163\,^{+0.065}_{-0.064}$ \\
CDF					&	$9.6$			& \cite{Aaltonen:2012ie}			&	$[-0.60,\, 0.12]$ at 68\% CL	&  $0.068 \pm 0.026 \pm 0.009$
\end{tabular}
\end{table}

This paper is organised as follows. Section~\ref{sec:pheno} presents the phenomenological aspects related to the measurement. 
Section~\ref{sec:detector} presents the LHCb detector. In Sect.~\ref{sec:selection} the selection of $\BtoJpsiKK$ candidates is described.
Section~\ref{sec:ptresolution} deals with decay time resolution, Sect.~\ref{sec:acceptance} with the decay time and angular acceptance 
effects and Sect.~\ref{sec:tagging} with flavour tagging. The maximum likelihood fit
is explained in Sect.~\ref{sec:fitting}. The results and systematic uncertainties for the $\BtoJpsiKK$ channel are given in Sections~\ref{sec:results} and \ref{sec:systematics}, 
the results for the $\BtoJpsipipi$ channel are given in Sect.~\ref{sec:resultspipi} and finally the combined results are presented in Sect.~\ref{sec:resultscombined}. Charge conjugation is implied throughout the paper.
%
%

\newcommand{\coefcosh}{\ensuremath{a_k}} \newcommand{\coefsinh}{\ensuremath{b_k}}
\newcommand{\coefcos}{\ensuremath{c_k}} \newcommand{\coefsin}{\ensuremath{d_k}}

\section{Phenomenology \label{sec:pheno}}
\begin{figure}[t]
\begin{center}
\begin{overpic}[scale=1, clip=true, trim=32mm 180mm 0mm 50mm]{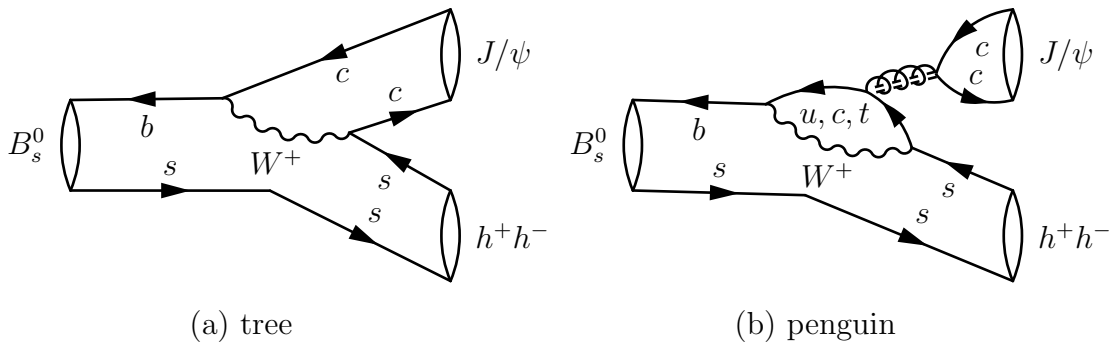}
\put(18,0){(a) tree}
\put(57,0){(b) penguin}
\end{overpic}
\end{center}\vspace{-0.2cm}
\caption{\small Feynman diagrams contributing to the decay $\Bs\to J/\psi h^{+}h^{-}$ within the SM, where $h = \pi, K$.}
\label{fig:jpsiphiDiag}
\end{figure}
The $\BtoJpsiKK$  decay proceeds predominantly via $ \BsToJPsiPhi$ with the $\phi$ meson subsequently decaying to $\Kp\Km$. 
In this case there are two intermediate vector particles and the $\Kp\Km$ pair is in a P-wave configuration. 
The final state is then a superposition of \CP-even and \CP-odd states depending upon the relative orbital angular momentum
of the \jpsi and the $\phi$.  The phenomenological aspects of this process are described in many articles, e.g., Refs.~\cite{Dighe:1995pd, Adeva:2009ny}. The main Feynman diagrams for $\BtoJpsiKK$ decays are shown in Fig.~\ref{fig:jpsiphiDiag}. 
The effects induced by the sub-leading penguin contributions are discussed, e.g., in Ref.~\cite{Faller:2008gt}. 
The same final state can also be produced with $\Kp\Km$ pairs in 
an S-wave configuration~\cite{Stone:2008ak}. This S-wave final state is \CP-odd.  
The measurement of $\phis$ requires the \CP-even and \CP-odd components to be disentangled by 
analysing the distribution of the reconstructed decay angles of the final-state particles.

\begin{figure}[t]
  \centering
  \includegraphics[scale=1, clip=true, trim=30mm 224.5mm 0mm 32.5mm]{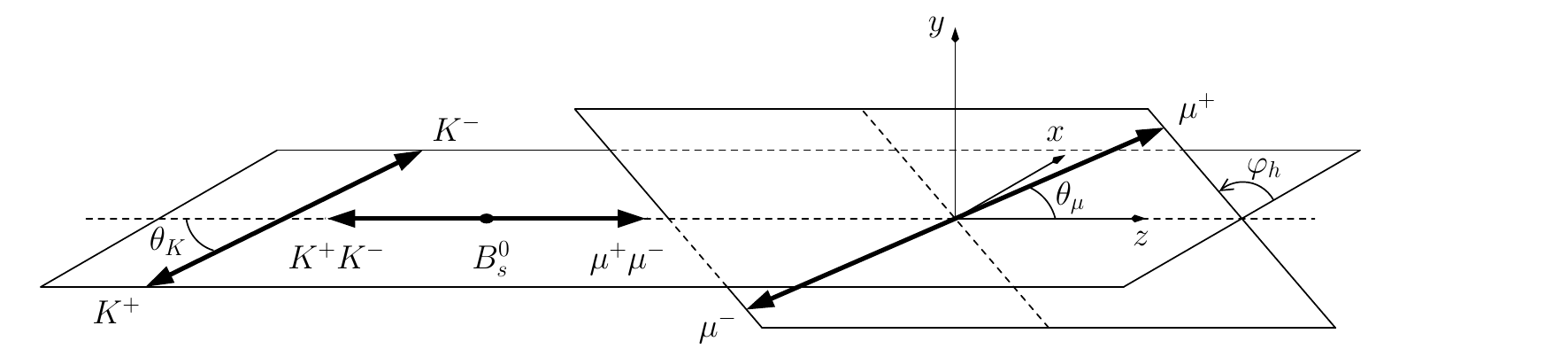}
  \caption{\small Definition of helicity angles as discussed in the text.}
  \label{fig:helicity}
\end{figure}

In contrast to Ref.~\cite{LHCb:2011aa}, this analysis uses the decay angles defined in the helicity basis as this simplifies the angular description of the background and acceptance. 
The helicity angles are denoted by $\Omega = (\cos\thetaK, \cos\thetamu, \phihel)$ and their definition is shown in Fig.~\ref{fig:helicity}. 
The polar angle \thetaK{} (\thetamu{}) is the angle between the $\Kp$ ($\mup$) momentum and the direction opposite to the 
\Bs{} momentum in the $\Kp\Km$ ($\mup\mun$) centre-of-mass system.
The azimuthal angle between the $\Kp\Km$ and $\mup\mun$ decay planes is \phihel. This angle is
defined by a rotation from the $\Km$ side of the $\Kp\Km$ plane to the $\mup$ side of the $\mup\mun$ plane. The rotation is positive in
the $\mup\mun$ direction in the \Bs{} rest frame. A definition of the angles in terms of the particle momenta is given in Appendix~\ref{app:angles}.

The decay can be decomposed into four time-dependent complex amplitudes, $A_i(t)$. 
Three of these arise in the P-wave decay and correspond to the relative orientation of the linear polarisation vectors of the $\jpsi$ and  $\phi$ mesons, 
where $i \in \{0, \parallel, \perp\}$ and refers to the longitudinal, transverse-parallel and transverse-perpendicular orientations, respectively. The single $\Kp\Km$ \mbox{S-wave} amplitude is denoted by $A_{\rm S}(t)$.

The distribution of the decay time and angles for a \Bs\ meson produced at time $t=0$ is described by a
sum of ten terms, corresponding to the four polarisation amplitudes
and their interference terms. Each of these is given by the product of a
time-dependent function and an angular function~\cite{Dighe:1995pd}
\begin{equation}
  \frac{\deriv^{4} \Gamma(\Bs\to J/\psi K^{+}K^{-}) }{\deriv t \;\deriv\Omega} \; \propto \;
  \sum^{10}_{k=1} \: h_k(t) \: f_k( \Omega) \,.
  \label{eq:Eqbsrate}
\end{equation}
The time-dependent functions $h_k(t)$ can be written as
\begin{multline}
  h_k (t) \; = \; N_k e^{- \Gs t} \: [ 
  \coefcosh \cosh\left(\tfrac{1}{2} \DGs t\right)  
    + \coefsinh \sinh\left( \tfrac{1}{2} \DGs t\right) \\ 
    + \coefcos \cos(\dms t) \, + \coefsin \sin(\dms t)],
  \label{eq:timefunc}
\end{multline}
where $\dms{}$ is the mass difference between the heavy and light \Bs\
mass eigenstates.
The expressions for the $f_k( \Omega)$ and the coefficients of  Eq.~\ref{eq:timefunc} are given in 
Table~\ref{tab:functions}~\cite{Dunietz:2000cr, Xie:2009fs}. 
The coefficients $N_k$ 
are expressed in terms of the $A_i(t)$ at $t=0$, from now on denoted as $A_i$. 
The amplitudes are parameterised by
$|A_i|e^{i\delta_{i}}$ with the conventions $\delta_0=0$ and 
$\azerosq + \aparsq + \aperpsq  = 1$. The S-wave fraction is defined as
\mbox{$\fS = |A_{\rm S}|^{2} / (  \azerosq + \aparsq + \aperpsq +  |A_{\rm S}|^{2}) = |A_{\rm S}|^{2} / (
 |A_{\rm S}|^{2} + 1 )$.}

\begin{table}[t]
\caption{\small Definition of angular and time-dependent functions.\label{tab:functions}}
\scriptsize
\newcommand{\Dterm}{D}
\newcommand{\Sterm}{S}
\newcommand{\Cterm}{C}
\[
\begin{array}{c|c|c|c|c|c|c}
  k  & f_k(\thetamu,\thetaK, \phihel) & N_k                & \coefcosh                  & \coefsinh & \coefcos & \coefsin \\
    \hline
  \rule{0mm}{3mm} 1  & 2\cos^2\thetaK \sin^2\thetamu  						& |A_0|^2         		& 1                         		 		& \Dterm 						& \Cterm 				& -\Sterm \\
  2  &  \sin^2\thetaK \left(1 - \sin^2\thetamu \cos^2\phihel\right)  	& |A_\||^2         		& 1                          				& \Dterm 						& \Cterm 				& -\Sterm \\
  3  &  \sin^2\thetaK \left(1 - \sin^2\thetamu \sin^2\phihel\right)  		& |A_\perp|^2      		& 1                          				& -\Dterm 						& \Cterm 				& \Sterm  \\
  4  & \sin^2\thetaK \sin^2\thetamu \sin2\phihel                     		& |A_\|A_\perp| 	& \Cterm  \sin(\delperp-\delpar)  	& \Sterm\cos(\delperp-\delpar) 		& \sin(\delperp-\delpar) 	& \Dterm \cos(\delperp-\delpar) \\
  5  & \tfrac{1}{2}\sqrt{2}\sin2\thetaK \sin2\thetamu \cos\phihel 		& |A_0 A_\||   		& \cos(\delpar - \delzero)   		&  \Dterm\cos(\delpar - \delzero) 	& \Cterm \cos(\delpar - \delzero)	& -\Sterm\cos(\delpar - \delzero)  \\
  6  & -\frac{1}{2}\sqrt{2} \sin2\thetaK \sin2\thetamu \sin\phihel 		& |A_0 A_\perp| 	& \Cterm \sin(\delperp - \delzero)	& \Sterm\cos(\delperp - \delzero)  	& \sin(\delperp - \delzero) &  \Dterm\cos(\delperp - \delzero) \\
  7  &  \tfrac{2}{3}\sin^2\thetamu            						& |A_{\rm S}|^2         		& 1                          				& -\Dterm 						& \Cterm				& \Sterm  \\
  8  & \tfrac{1}{3}\sqrt{6} \sin\thetaK \sin2\thetamu \cos\phihel 		& |A_{\rm S} A_\||   		& \Cterm    \cos(\delpar - \dels)   	& \Sterm \sin(\delpar - \dels) 		& \cos(\delpar - \dels) 	& \Dterm \sin(\delpar - \dels)  \\
  9  & -\tfrac{1}{3}\sqrt{6}\sin\thetaK \sin2\thetamu \sin\phihel 		& |A_{\rm S} A_\perp| 	& \sin(\delperp - \dels) 			& -\Dterm \sin(\delperp - \dels) 		& \Cterm 	\sin(\delperp - \dels)	& \Sterm\sin(\delperp - \dels) \\
  10 & \tfrac{4}{3}\sqrt{3}\cos\thetaK \sin^2\thetamu 				& |A_{\rm S} A_0|    		& \Cterm   \cos(\delzero - \dels)   	& \Sterm\sin(\delzero - \dels) 		& \cos(\delzero - \dels) 	&  \Dterm\sin(\delzero - \dels) \\
  
\end{array}
\]
\end{table}

For the coefficients $a_k,\ldots,d_k$,  three \CP{} violating observables are introduced
\begin{equation} 
  C \; \equiv \; \frac{1 - |\lambda|^2}{1 + |\lambda|^2} \;, \qquad
  S \; \equiv \; \frac{2 \Im(\lambda) }{1 + |\lambda|^2} \;, \qquad 
  D \; \equiv \; -\frac{2 \Re(\lambda)}{1 + |\lambda|^2} \; ,
  \label{eqn:SDC1}
\end{equation}
\noindent where the parameter $\lambda$ is defined below. These definitions for $S$ and $C$ correspond to those adopted by HFAG~\cite{Amhis:2012bh}
and the sign of $D$ is chosen such that it is equivalent to the symbol 
$A^{\Delta \Gamma}_f$ used in Ref.~\cite{Amhis:2012bh}.
The \CP-violating phase $\phis$ is  defined by  $\phis \equiv -\arg( \lambda )$ and hence $S$ and $D$ can be written as 
\begin{equation}
  S \; \equiv \; -\frac{2  |\lambda| \sin{\phis} }{1 + |\lambda|^2} \;, \qquad 
  D \; \equiv \; -\frac{2 |\lambda| \cos{\phis} }{1 + |\lambda|^2} \; .
\end{equation}
\noindent The parameter $\lambda$ 
describes \CP{} violation in the interference between mixing and decay, and
is derived from the \CP{}-violating parameter~\cite{Branco:1999fs} associated with  
each polarisation state~$i$ 
\begin{equation}
  \lambda_i \; \equiv \; \frac{q}{p} \; \frac{\bar{A}_i}{A_i},
\end{equation}
where $A_i$ ($\bar{A}_i$) is the amplitude for a $B_s^{0}$
($\Bsb$) meson to decay to final state $i$ and the complex parameters $p =
\braket{B_s^{0}}{B_L}$ and $q=\braket{\Bsb}{B_L}$ describe the
relation between mass and flavour eigenstates.
The polarisation states $i$ have \CP{} eigenvalue
$\eta_i =  +1   \:\  \text{for $i\in\{0,\parallel\}$}$ and $\eta_i =-1 \: \text{for $i\in\{\perp,{\rm S}\}$}$.
Assuming that any possible \CP{} violation in the
decay is the same for all amplitudes, then the product $\eta_i\bar{A}_i/ A_i$ is independent of $i$. 
The polarisation-independent \CP{}-violating parameter $\lambda$ is then defined such that $\lambda_i= \eta_i \lambda$. 
The differential decay rate for a \Bsb meson produced at time $t=0$
can be obtained by changing the sign of \coefcos{} and \coefsin{} 
and by including a relative factor $|p/q|^2$. 

The expressions are invariant under the transformation
\begin{equation}
  (\phis,\DGs, \delta_0,\delpar,\delperp,\dels) \longmapsto
  (\pi-\phis,-\DGs, -\delta_0,-\delpar,\pi-\delperp,-\dels) \: ,
\end{equation}
which gives rise to a two-fold ambiguity in the results.

In the selected $\pip\pim$ invariant mass range the \CP-odd fraction of
$\BtoJpsipipi$ decays is greater than 97.7\% at 95\% confidence level (CL) as described
in Ref.~\cite{LHCb:2012ae}.  As a consequence, no angular
analysis of the decay products is required and the differential decay rate can be simplified to
\begin{equation}
  \frac{\deriv \Gamma(\BtoJpsipipi) }{\deriv t } \; \propto \;  h_7(t).
  \label{eq:Eqbsrate-pipi}
\end{equation}

\section{Detector}
\label{sec:detector}

The \lhcb detector~\cite{Alves:2008zz} is a single-arm forward
spectrometer covering the \mbox{pseudorapidity} range $2<\eta <5$, designed
for the study of particles containing \bquark or \cquark quarks. The
detector includes a high precision tracking system consisting of a
silicon-strip vertex detector surrounding the $pp$ interaction region,
a large-area silicon-strip detector located upstream of a dipole
magnet with a bending power of about $4{\rm\,Tm}$, and three stations
of silicon-strip detectors and straw drift-tubes placed
downstream. The combined tracking system has momentum resolution
$\Delta p/p$ that varies from 0.4\% at 5\gevc to 0.6\% at 100\gevc,
and impact parameter resolution of 20\mum for tracks with high
transverse momentum. Charged hadrons are identified using two
ring-imaging Cherenkov detectors~\cite{2012arXiv1211.6759A}.
Photon, electron and hadron
candidates are identified by a calorimeter system consisting of
scintillating-pad and pre-shower detectors, an electromagnetic
calorimeter and a hadronic calorimeter. Muons are identified by a
system composed of alternating layers of iron and multiwire
proportional chambers. The trigger consists of a hardware stage, based
on information from the calorimeter and muon systems, followed by a
software stage which applies a full event reconstruction~\cite{Aaij:2012me}.

Simulated $pp$ collisions are generated using
\pythia~6.4~\cite{Sjostrand:2006za} with a specific \lhcb
configuration~\cite{LHCb-PROC-2010-056}.  Decays of hadronic particles
are described by \evtgen~\cite{Lange:2001uf} in which final state
radiation is generated using \photos~\cite{Golonka:2005pn}. The
interaction of the generated particles with the detector and its
response are implemented using the \geant
toolkit~\cite{Allison:2006ve, *Agostinelli:2002hh} as described in
Ref.~\cite{LHCb-PROC-2011-006}.

%
%
\newcommand{\BsJphi}{\decay{\Bs}{\Jpsi\phi}}             
\newcommand{\Jpsi}{\ensuremath{J\!/\!\psi}}

\section{\boldmath Selection of $\BtoJpsiKK$ candidates \label{sec:selection}}

The reconstruction of $\BtoJpsiKK$ candidates proceeds using the decays 
\mbox{$\Jpsi\to\mu^{+}\mu^{-}$}
combined with a pair of oppositely charged kaons. 
Events are first required to pass a hardware trigger~\cite{Aaij:2012me}, which selects events containing muon or hadron candidates
with high transverse momentum ($p_{\rm T}$). 
The subsequent software trigger~\cite{Aaij:2012me} is composed of two stages, the first of which
performs a partial event reconstruction. 
Two types of first-stage software trigger are employed. For the first type, events are required to have two well-identified oppositely-charged muons with invariant mass
larger than $2.7\gevcc$. 
 This trigger has an almost uniform acceptance as a function of decay time and will be referred to as {\it unbiased}.
 For the second type there must be at least one muon (one high-$\pt$ track) with 
 transverse momentum larger than $1\gevc$ ($1.7\gevc$) and impact parameter larger than 100\mum with respect to the PV. 
This trigger introduces a non-trivial acceptance as a function of decay time and will be referred to as {\it biased}.
The second stage of the trigger performs a full event reconstruction and only retains events containing a $\mu^{+}\mu^{-}$ 
pair with invariant mass within $120\mevcc$ of the \Jpsi\ mass~\cite{pdg} and which form
a vertex that is significantly displaced from the PV, introducing another small decay time biasing effect. 

The final \Bs candidate selection is performed by applying kinematic and particle identification criteria 
to the final-state tracks. The \Jpsi~meson candidates are formed from two oppositely-charged 
particles, originating from a common vertex, which have been identified as muons and which have 
$p_{\rm T}$ larger than 500\mevc. 
The invariant mass of the $\mu^{+}\mu^{-}$ pair, $m(\mu^+\mu^-)$, 
must be in the range $[3030, 3150]\mevcc$. 
During subsequent steps of the selection,  $m(\mu^+\mu^-)$ is constrained to the 
\Jpsi\ mass~\cite{pdg}.

The $\Kp\Km$ candidates are formed from two oppositely-charged particles that have been identified as
kaons and which originate from a common vertex. The $\Kp\Km$ pair  is required to have a $p_{\rm T}$ larger than 1\gevc.
The invariant mass of the $\Kp\Km$ pair, $m(\Kp\Km)$, must be in the 
range $[990, 1050]\mevcc$. 

The \Bs candidates are reconstructed by combining the \Jpsi\ candidate with the $\Kp\Km$ pair, requiring their 
invariant mass $m(\Jpsi\Kp\Km)$ to be in the range $[5200, 5550]\mevcc$. The decay time, $t$,
of the \Bs\ candidate is calculated from a vertex and kinematic fit that constrains the $\Bs\to\Jpsi K^{+}K^{-}$ 
candidate to originate from its associated PV~\cite{Hulsbergen:2005pu}. The $\chi^{2}$ of the 
fit (which has 7 degrees of freedom) is required to be less than 35. Multiple \Bs candidates are 
found in less than $1\%$ of events; in these cases the candidate with the smallest
$\chi^{2}$ is chosen. \Bs candidates are required to have decay time in the range $[0.3, 14.0]\ps$; the lower
bound on the decay time suppresses a large fraction of the prompt combinatorial background 
whilst having a negligible effect on the sensitivity to \phis. The kinematic fit evaluates an estimated decay time uncertainty, $\sigma_{t}$.
Candidates with $\sigma_{t}$ larger than 0.12\ps are removed from the event sample. 

\begin{figure}[t]
\centering
  \includegraphics[trim=13mm 11mm 8mm 25mm, clip=true, width=0.5\textwidth]{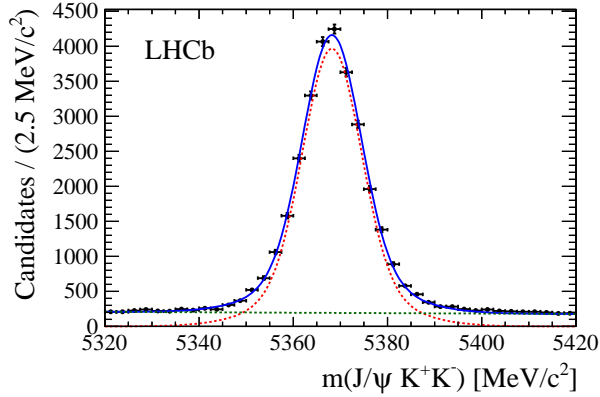}
  \caption{\small
  Invariant mass distribution of the selected $\Bs\to\Jpsi K^{+}K^{-}$ candidates. The mass of the $\mu^{+}\mu^{-}$ pair is constrained to the  \Jpsi\ mass~\cite{pdg}. 
  Curves for the fitted contributions from signal (dotted red), background (dotted green) and their combination (solid blue) are overlaid.
}
 \label{fig:mass}
\end{figure}

\begin{figure}[t]
\centering
		\begin{overpic}[trim=13mm 11mm 8mm 25mm, clip=true, width=0.49\textwidth]{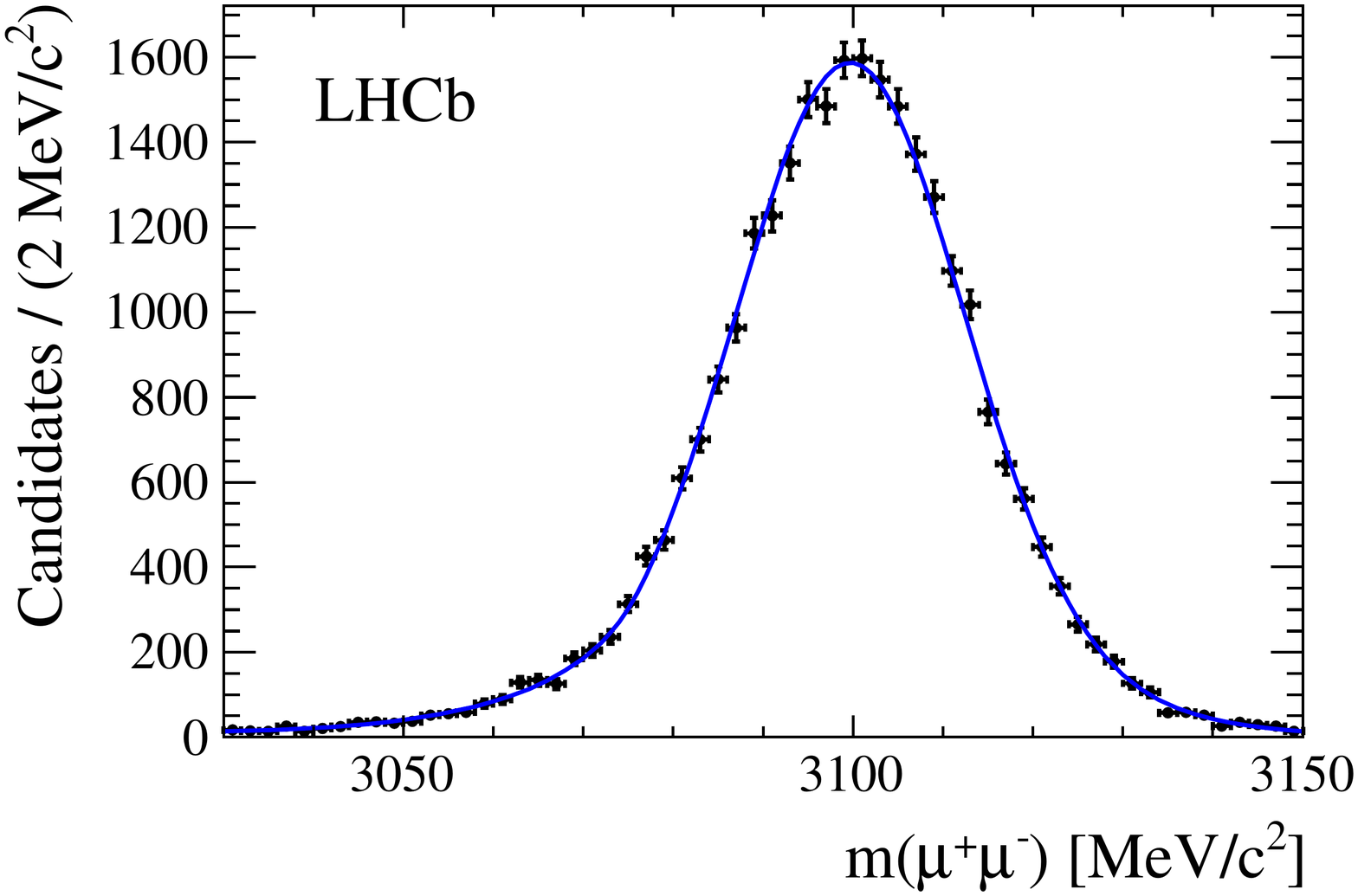}
		\put(26,46){(a)}
		\end{overpic}
                \hspace*{0.00\textwidth}
		\begin{overpic}[trim=13mm 11mm 8mm 25mm, clip=true, width=0.49\textwidth]{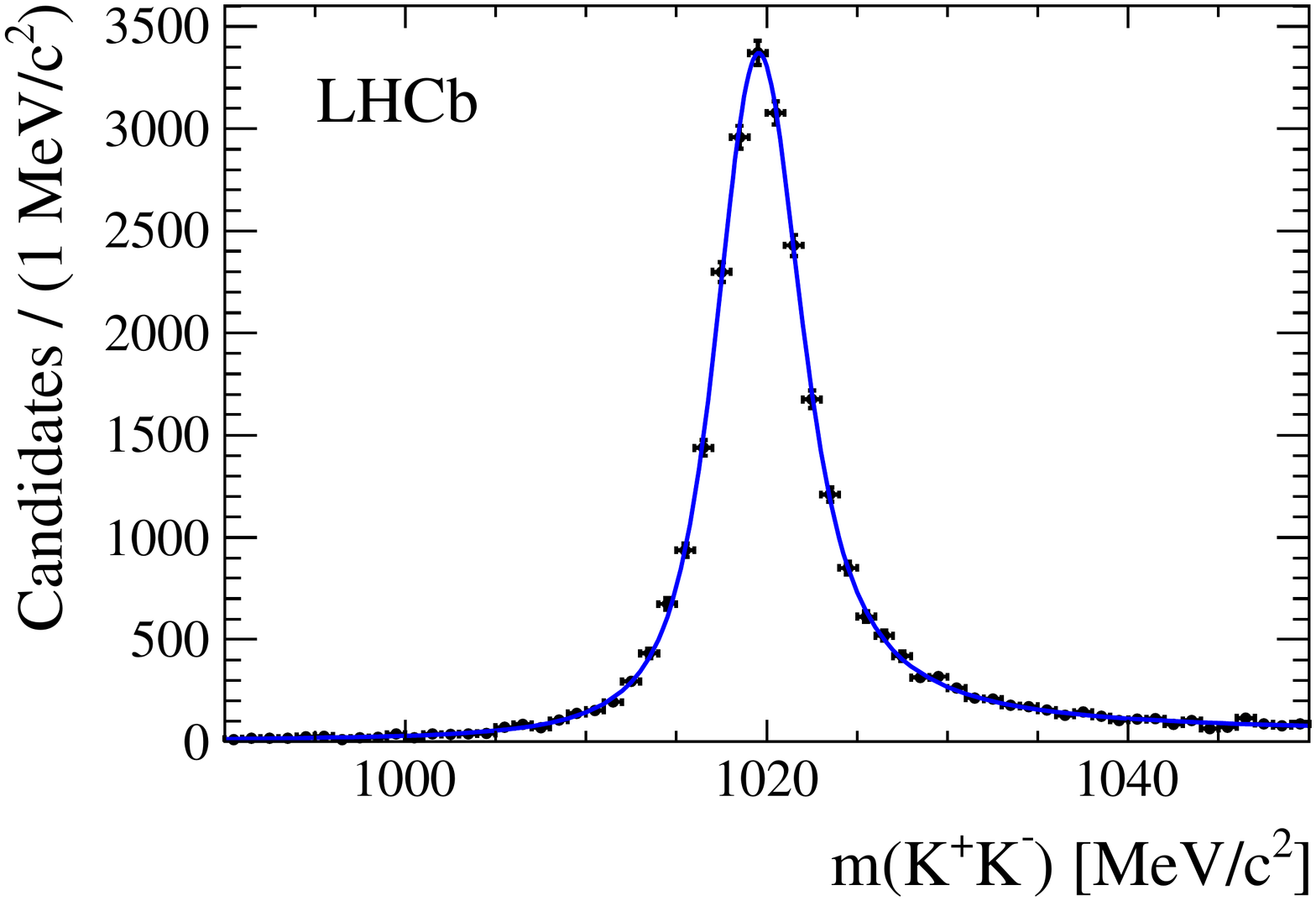}
		\put(26,46){(b)}
		\end{overpic}
 \caption{\small Background subtracted invariant mass distributions of the (a) $\mu^{+}\mu^{-}$ and (b) $K^{+}K^{-}$ systems in the selected sample of $\Bs\to\Jpsi K^{+}K^{-}$ candidates. 
 The solid blue line represents the fit to the data points described in the text.}
\label{fig:jpsi_phi_mass}
\end{figure}

Figure~\ref{fig:mass} shows the $m(\Jpsi\Kp\Km)$ distribution for
events originating from both the unbiased and biased triggers, along with corresponding projection of an unbinned 
maximum log-likelihood fit to the sample. The probability density function (PDF) used for the fit is composed
of the sum of two Gaussian functions with a common mean and separate widths and an exponential function for
the combinatorial background. In total, after the trigger and full offline selection requirements, there are
$27\,617 \pm 115$ $\Bs\to\Jpsi K^{+}K^{-}$ signal events found by the fit. 
Of these, $23\,502 \pm 107$ were selected by the unbiased trigger and $4115 \pm 43$ were exclusively selected by the biased trigger. The uncertainties quoted here come from propagating the 
uncertainty on the signal fraction evaluated by the fit.

Figure~\ref{fig:jpsi_phi_mass} shows the invariant mass of the $\mu^{+}\mu^{-}$ and $K^{+}K^{-}$ pairs satisfying
the selection requirements. The background has been subtracted using the \sPlot~\cite{splot}
technique with $m(\Jpsi\Kp\Km)$ as the discriminating variable.
In both cases fits are also shown. For the di-muon system the fit model is a double Crystal Ball shape~\cite{Skwarnicki:1986xj}. For 
the di-kaon system the total fit model is the sum of a relativistic P-wave Breit-Wigner distribution convolved with a Gaussian function to model the dominant $\phi$
meson peak and a polynomial function to describe the small $K^{+}K^{-}$ S-wave component.

%
%

\section{Decay time resolution \label{sec:ptresolution}}

If the decay time resolution is not negligibly small compared to the
\Bs\ meson oscillation period $2\pi / \dms \approx 350$~fs, it affects the
measurement of the oscillation amplitude, and thereby $\phi_s$. For a
given decay time resolution, $\sigma_t$, the dilution of the amplitude can be
expressed as ${\cal D} = \exp( - \sigma_t^2 \dms^2/2)$~\cite{Moser:1996xf}. The
relative systematic uncertainty on the dilution directly translates
into a relative systematic uncertainty on $\phi_s$.

\begin{figure}[t]
  \centerline{
  \includegraphics[width=0.50\textwidth, height=5.2cm]{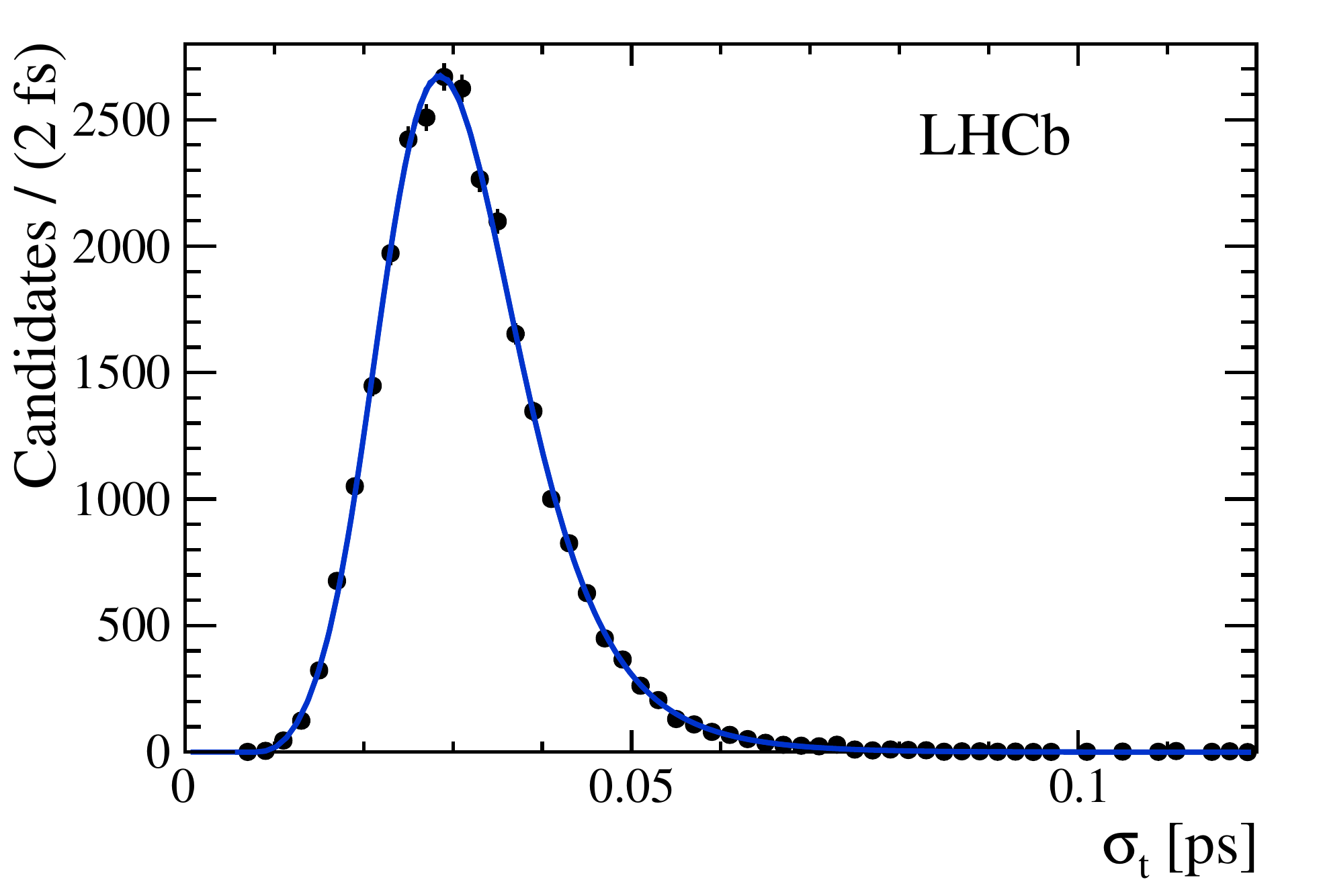}
  }
  \caption{\small
    Decay time resolution, $\sigma_{t}$, for selected $\Bs\to\Jpsi K^{+}K^{-}$ signal events. 
    The curve shows a fit to the data of the sum of two gamma distributions with a common mean. 
 }
\label{fig:time_err}
\end{figure}

For each reconstructed candidate, $\sigma_t$ is
estimated by the vertex fit with which the decay time is
calculated. The signal distribution of $\sigma_t$ is 
shown in Fig.~\ref{fig:time_err} where the \sPlot\
technique is used to subtract the background. To
account for the fact that track parameter resolutions are not
perfectly calibrated and that the resolution function is not Gaussian,
a triple Gaussian resolution model is constructed
\begin{equation}
  R(t; \sigma_t) \; = \; \sum_{i=1}^3 \: \frac{f_i}{\sqrt{2\pi}
   r_i  \sigma_t} \: \exp\left[ -
    \frac{( t - d )^2}{ 2 r_i^2 \sigma_t^2} \right],
    \label{eqn:pe_reso}
\end{equation}
where $d$ is a common small offset of a few fs,
$r_i$ are event-independent resolution scale factors and $f_i$ is the
fraction of each Gaussian component, normalised such that $\sum f_i =
1$.

\begin{figure}[t]
  \centerline{
  \begin{overpic}[width=0.50\textwidth]{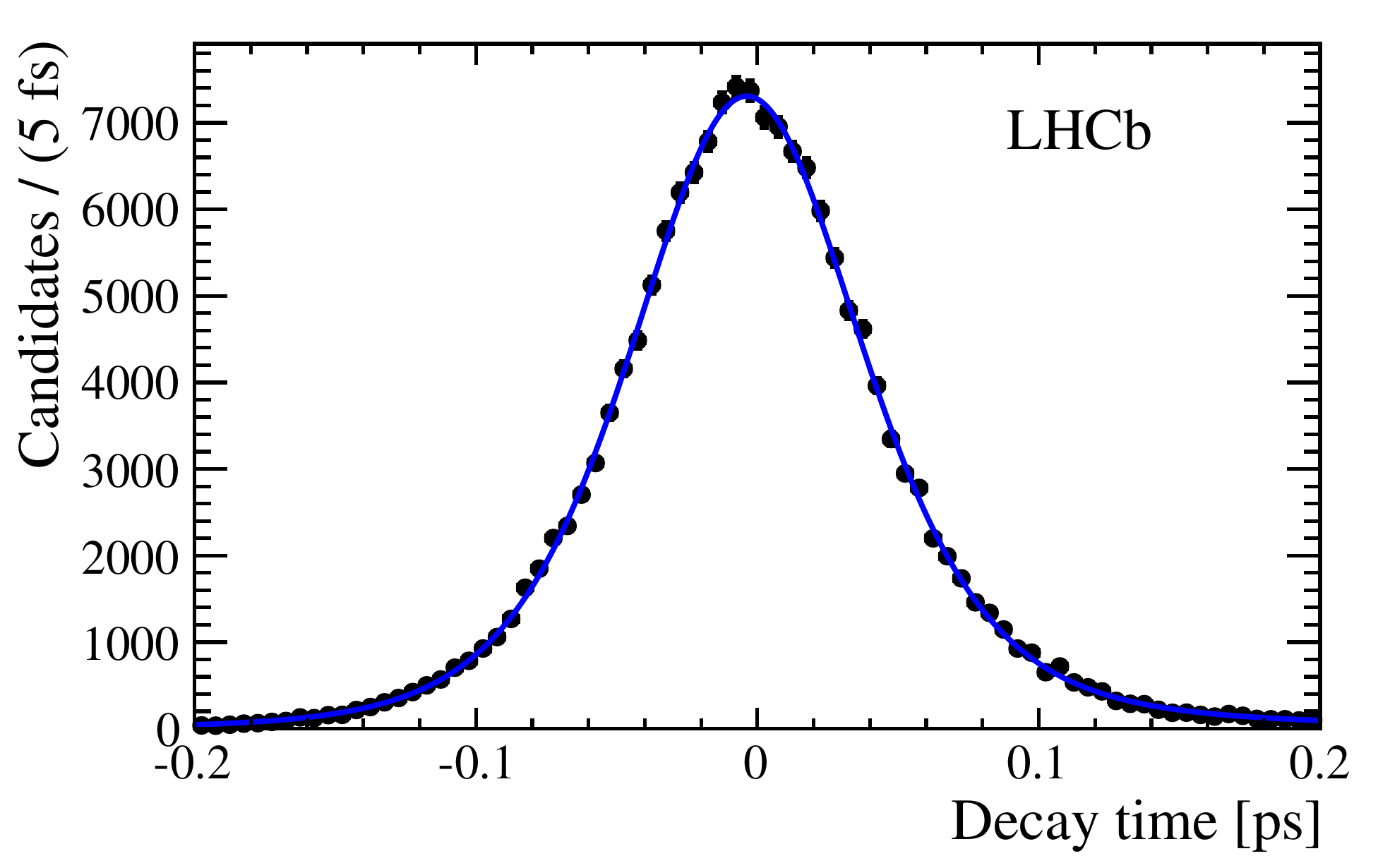}
  \end{overpic}
    \begin{overpic}[width=0.50\textwidth]{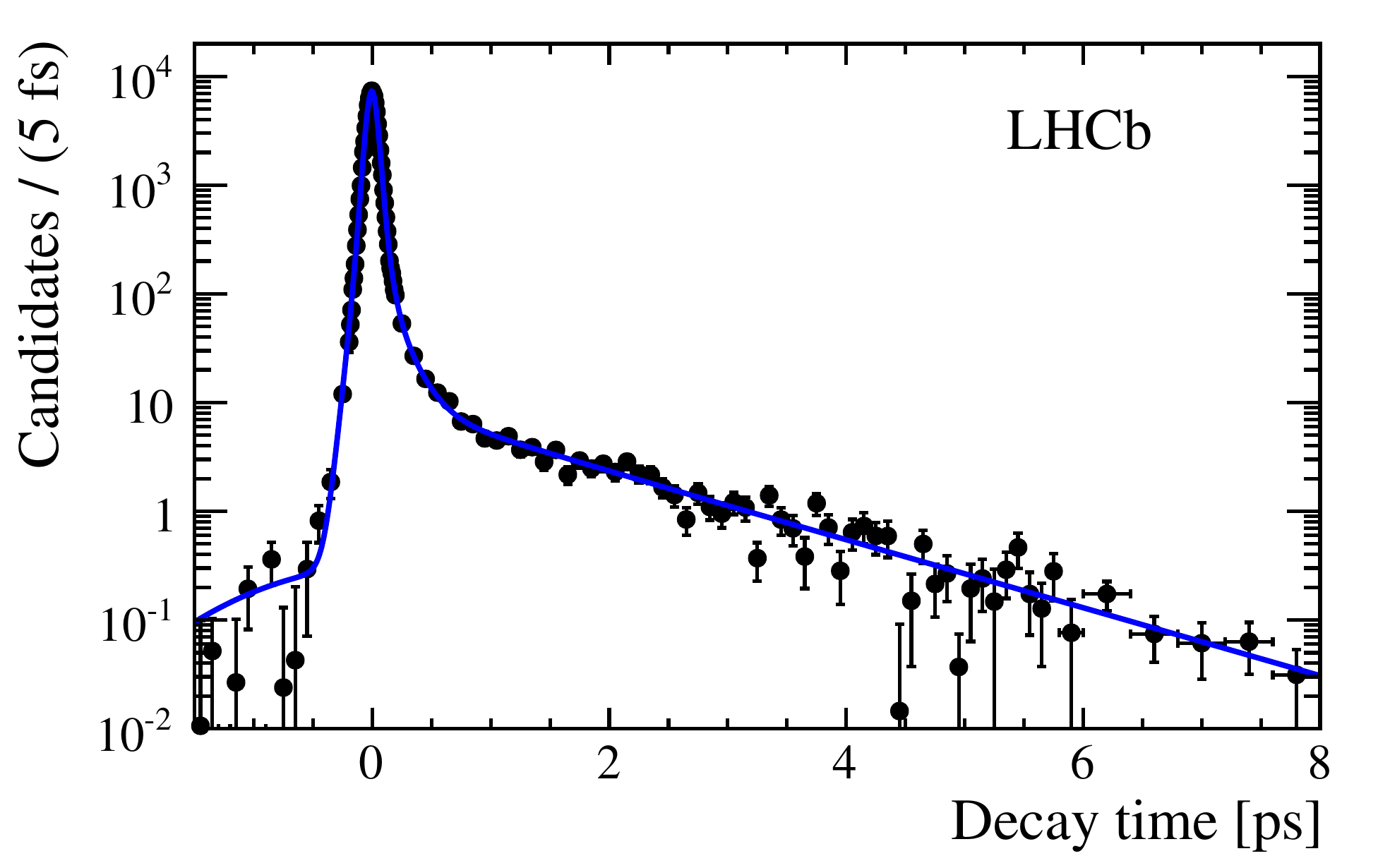}
  \end{overpic}
  }
  \caption{\small
	Decay time distribution of prompt $\jpsi\Kp\Km$ candidates. 
    The curve (solid blue) is the decay time model convolved with a Gaussian
    resolution model. The decay time model consists of a delta function for the
    prompt component and two exponential functions with different decay
    constants, which represent the $\Bs\to\Jpsi K^{+}K^{-}$ signal and long-lived background, respectively.
    The decay constants are determined from the fit. The same dataset is shown in both plots, on different scales.}
\label{fig:resolution}
\end{figure}

The scale factors are estimated from a sample of prompt $\mu^+ \mu^-
K^+ K^-$ combinations that pass the same selection criteria as the
signal except for those that affect the decay time distribution. This sample consists
primarily of prompt combinations that have a true decay time of
zero. Consequently, the shape of the decay time distribution close to
zero is representative of the resolution function itself.

Prompt combinations for which the muon pair
originates from a real $\jpsi$ meson have a better resolution than those
with random muon pairs. Furthermore, fully simulated events confirm 
that the resolution evaluated using prompt
$\jpsi\to\mu^+\mu^-$ decays with two random kaons is more representative for
the resolution of $B_s^{0}$ signal decays than the purely combinatorial
background.  Consequently, in the data only $\jpsi K^+
K^-$ events are used to estimate the resolution function. These are
isolated using the \sPlot\ method to subtract the 
$\mu^+\mu^-$ combinatorial background.

The background subtracted decay time distribution for $\jpsi K^+ K^-$
candidates is shown in Fig.~\ref{fig:resolution} using linear and logarithmic
scales. The
distribution is characterised by a prompt peak and a tail due to \jpsi{} mesons
from $B$ decays. The resolution model parameters are determined
by fitting the distribution with a decay time model that consists of a prompt peak
and two exponential functions, convolved with the resolution model given in
Eq.~\ref{eqn:pe_reso}.

The per-event resolution receives contributions both from the vertex
resolution and from the momentum resolution. The latter contribution
is proportional to the decay time and cannot be calibrated with the
prompt $\jpsi K^+ K^-$ control sample. When using a scale factor for
the resolution there is an assumption that the vertex contribution and
the momentum contribution have a common scale. This assumption is
tested in simulations and a systematic uncertainty is assigned.

The effective dilution of the resolution function is calculated by taking its
Fourier transform calculated at frequency $\dms$~\cite{Moser:1996xf}
\begin{equation}
  {\cal D} \; = \; \int_{-\infty}^\infty {\rm d}t \; \cos(\dms t) \: R(t ; \sigma_t).
\end{equation}
Taking into account the distribution of the per-event resolution, the
effective dilution for the calibrated resolution model is $0.72 \pm
0.02$. This dilution corresponds to an
effective single Gaussian resolution of approximately 45~fs. The
systematic uncertainty accounts for uncertainties due to the momentum
resolution scale and other differences between the control sample and
signal decays. It is derived from simulations.

The sample used to extract the physics parameters of interest consists
only of events with $t>0.3$~ps. The observed decay time distribution of
these events is not sensitive to details of the resolution
function. Therefore, in order to simplify the fit procedure the
resolution function for the final fit (described in Sect.~\ref{sec:fitting}) 
is modelled with a
single Gaussian distribution with a resolution scale factor, $\eventresscale$, chosen such that
its effective dilution corresponds to that of the multiple Gaussian
model. This scale factor is \mbox{$\eventresscale = 1.45 \pm 0.06$}. 

\section{Acceptance}
\label{sec:acceptance}
\newcommand{\eff}{\varepsilon(t,\Omega)}
\newcommand{\efft}{\varepsilon_t(t)}
\newcommand{\effa}{\varepsilon_\Omega}
\newcommand{\meff}{\langle\varepsilon\rangle}
\newcommand{\mefft}{\langle\varepsilon_t\rangle}
\newcommand{\meffa}{\langle\varepsilon_a\rangle}

There are two distinct decay time acceptance effects that influence the \Bs\ decay time distribution.  First, there is a decrease in
reconstruction efficiency for tracks with a large impact parameter with respect to the beam line. This effect is present both in the
trigger and the offline reconstruction, and translates to a decrease in the \Bs meson reconstruction efficiency as a function of its decay
time. This decrease is parameterised by a linear acceptance function $\efft\propto(1+\beta t)$, which multiplies the time dependent
$\Bs\to\Jpsi K^{+}K^{-}$ PDF described below. The parameterisation is determined using a control sample of $\B^{\pm}\to\Jpsi K^{\pm}$ events from 
data and simulated \BsJphi\ events, leading to \mbox{$\beta=(-8.3 \pm 4.0)\times10^{-3}\invps$}. 
The uncertainty directly translates to a $4.0\times 10^{-3}\invps$ systematic uncertainty on $\Gamma_{s}$.

\begin{figure}[tb]
  \centerline{
    \begin{overpic}[width=0.50\textwidth]{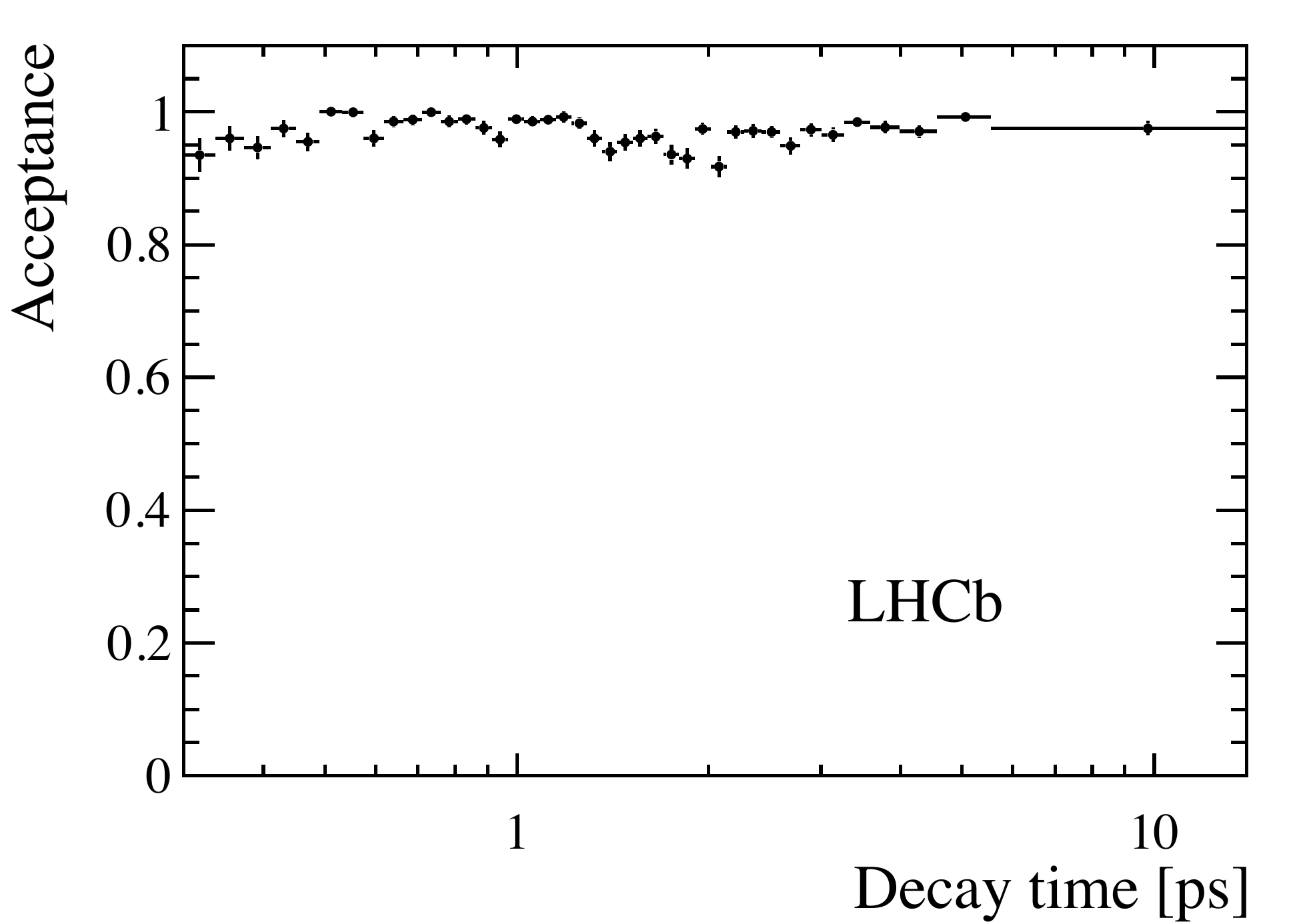}
  \put(66,34){(a)}
  \end{overpic}
      \begin{overpic}[width=0.50\textwidth]{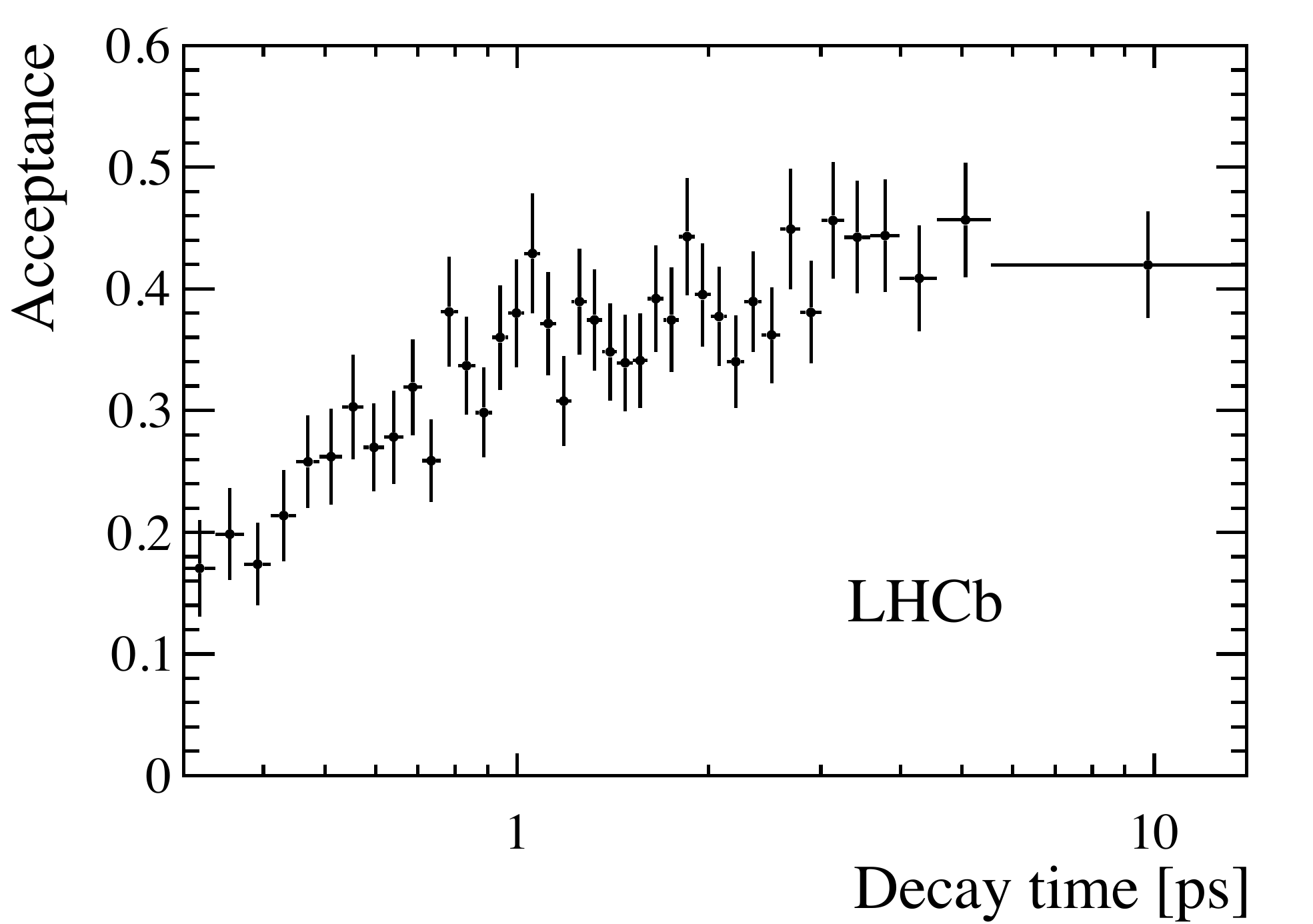}
  \put(66,34){(b)}
  \end{overpic}
  }
  \caption{\small\Bs decay time trigger-acceptance functions obtained from data. The unbiased trigger category is shown on (a) an absolute scale and (b) the biased trigger category on an arbitrary scale.}
  \label{fig:time_acceptance}
\end{figure}

Secondly, a non-trivial decay time acceptance is introduced by the trigger selection.  Binned functional descriptions of
the acceptance for the unbiased and biased triggers are obtained from the data by exploiting the sample of \Bs candidates that are also selected by a trigger that has no decay time bias, but was only used for a fraction of the recorded data.  Figure~\ref{fig:time_acceptance} shows the corresponding acceptance functions 
that are included in the fit described in Sect.~\ref{sec:fitting}.

\begin{figure}[tbp]
    \begin{overpic}[trim=45mm 13mm 19mm 32mm, clip=true, width=0.32\textwidth]{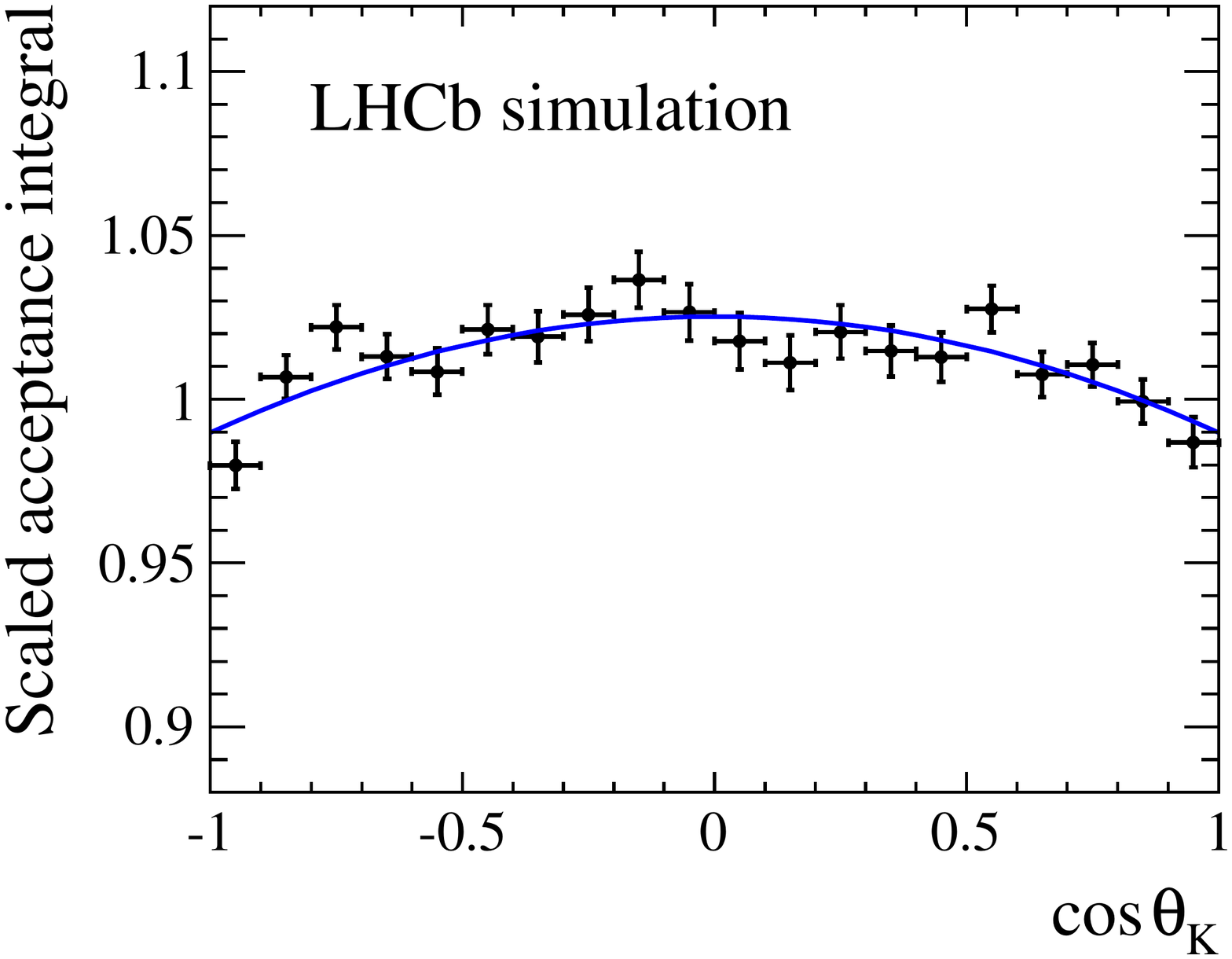}
  \put(70,67){\small (a)}
  \end{overpic}
  \hspace*{0.01\textwidth}%
  \begin{overpic}[trim=45mm 13mm 19mm 32mm, clip=true, width=0.32\textwidth]{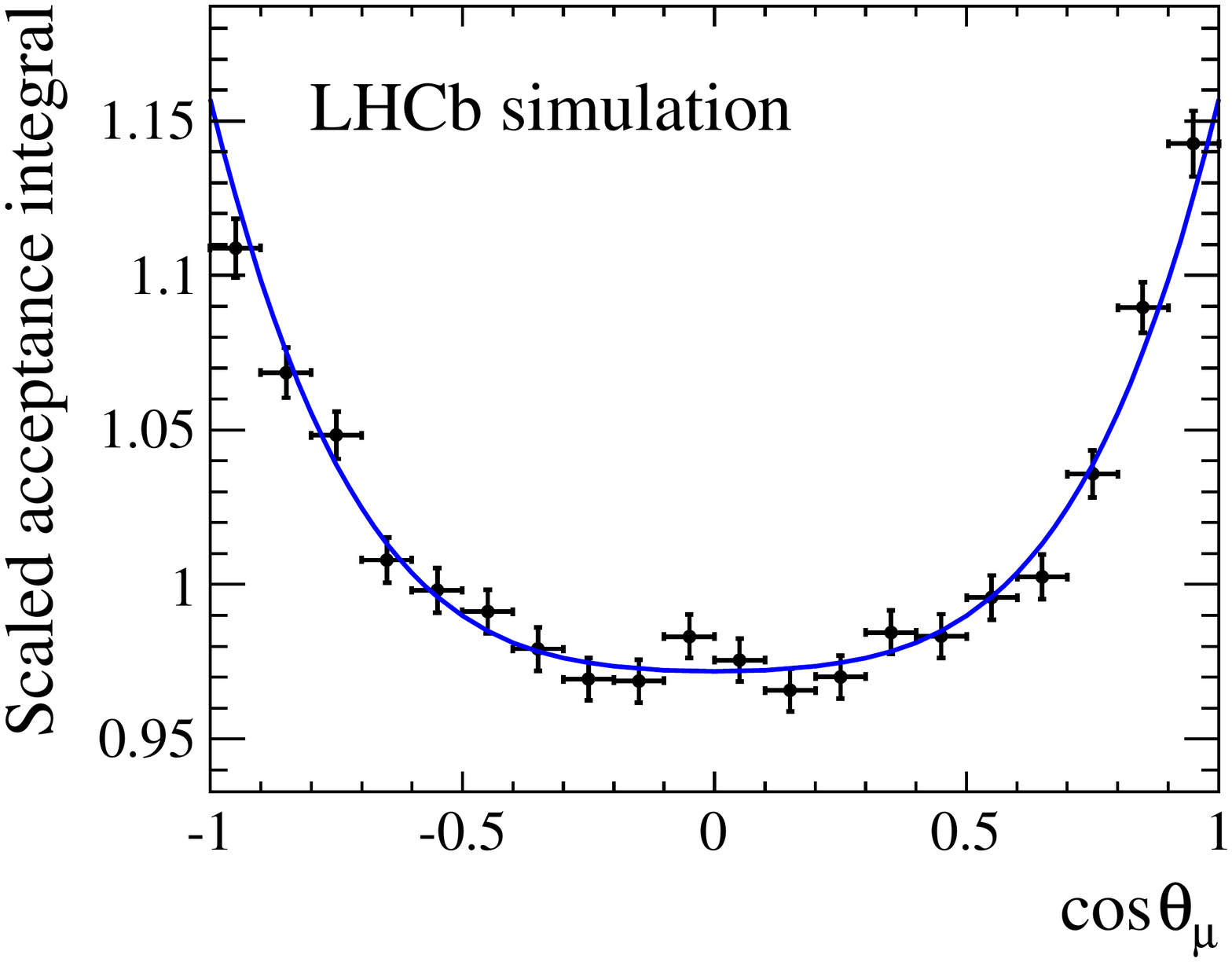}%
  \put(70,67){\small (b)}
  \end{overpic}
  \hspace*{0.01\textwidth}%
  \begin{overpic}[trim=45mm 13mm 19mm 32mm, clip=true, width=0.32\textwidth]{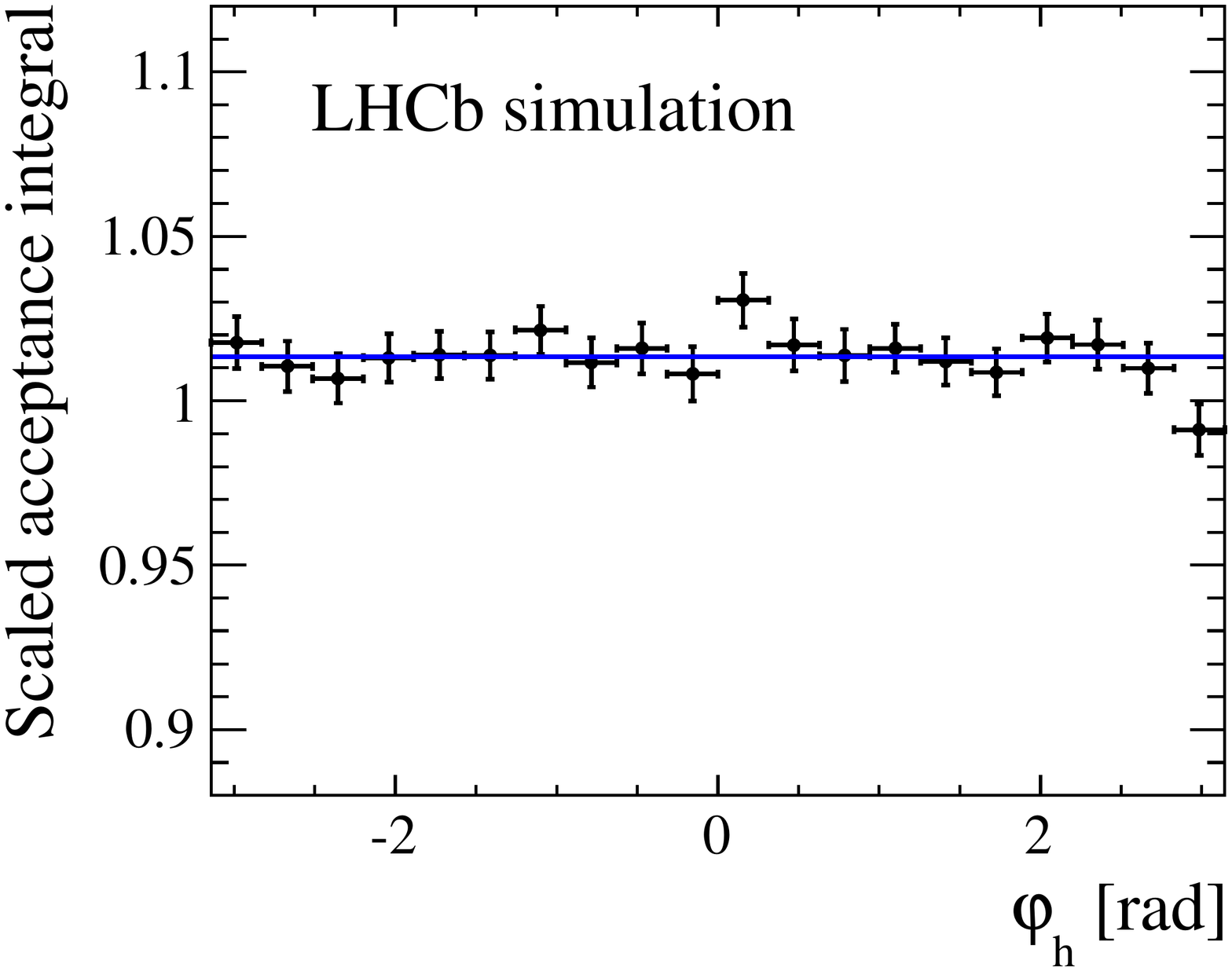}
  \put(70,67){\small (c)}
  \end{overpic}
  \caption{\small Angular acceptance function evaluated with simulated \BsJphi\ events, scaled by the mean acceptance. The acceptance is shown as 
  a function of (a) $\cos\thetaK$, (b) $\cos\thetamu$ and (c) $\phihel$, where in all cases the acceptance is integrated over the other two angles.
           The points are obtained by summing the inverse values of the underlying physics PDF for simulated events and
           the curves represent a polynomial parameterisation of the acceptance.}
  \label{fig:angEffFuncIntegrals}
\end{figure}

The acceptance as a function of the decay angles is not uniform due to the forward geometry of LHCb and the requirements placed upon the momenta of
the final-state particles. The three-dimensional acceptance function, $\effa$, is determined using simulated events which are subjected to the
same trigger and selection criteria as the data. 
Figure~\ref{fig:angEffFuncIntegrals} shows the angular efficiency as a function of each decay angle, integrated over the other angles.
The relative acceptances vary by up to 20\% peak-to-peak. The dominant effect in $\cos\theta_{\mu}$ is due to the 
$\pt$ cuts applied to the muons.

The acceptance is included in the unbinned maximum log-likelihood fitting procedure to signal weighted 
distributions (described in Sect.~\ref{sec:fitting}). Since only a PDF to describe the signal is required,
the acceptance function needs to be included only in the normalisation of the PDF through the ten integrals $\int\mathrm{d}\Omega\, \varepsilon_{\Omega}(\Omega)\, f_k(\Omega)$. The acceptance factors for each event $i$, $\varepsilon_{\Omega}(\Omega_i)$,  appear only 
as a constant sum of logarithms and may be ignored in the likelihood maximisation.
The ten integrals are determined from the fully simulated events using the procedure described in Ref.~\cite{TristansThesis}. 

%
%

\section{\boldmath Tagging the $\Bs$ flavour at production  \label{sec:tagging}}

Each reconstructed candidate is identified by flavour tagging algorithms as either a \Bs meson ($q=+1$) or a \Bsb
meson ($q=-1$) at production. If the algorithms are unable to make a decision, the candidate is untagged ($q=0$).

The tagging decision, $q$,  is based upon both opposite-side and same-side tagging algorithms.
The opposite-side (OS) tagger relies on the pair production of \bquark and \bquarkbar quarks and infers the flavour of 
the signal \Bs meson from identification of the flavour of the other  \bquark-hadron.
The OS tagger uses the charge of the lepton ($\mu$, $e$) from semileptonic $b$ decays, the charge of the kaon from the $b\to c\to s$ 
decay chain and the charge of the inclusive secondary vertex reconstructed from $b$-hadron decay products.
The same-side kaon (SSK) tagger exploits the hadronization process of the 
$\overline{b}$(\bquark) quark forming the signal $B_s^0$(\Bsb) meson. 
In events with a \Bs candidate, the fragmentation of a \bquarkbar quark can lead to an extra \squarkbar quark being available to form a hadron, often leading to a charged kaon. 
This kaon  is correlated to the signal \Bs in phase space and the sign of the charge identifies its initial flavour. 

The probability that the tagging determination is wrong (estimated wrong-tag probability, $\eta$) is 
based upon the output of a neural network trained on simulated events. It is subsequently calibrated with data in
order to relate it to the true wrong-tag probability of the event, $\omega$, as described below.

The tagging decision and estimated wrong-tag probability are used event-by-event 
in order to maximise the tagging power, \effD, which represents the effective reduction of the signal sample size due to imperfect tagging.
In this expression \etag is the tagging efficiency, i.e., the fraction of events that are assigned 
a non-zero value of $q$, and ${\cal D} = 1 - 2\omega$ is the dilution.

\subsection{Opposite side tagging\label{sec:OStagging}}
The OS tagging algorithms and the procedure used to optimise and calibrate them are described in Ref.~\cite{Aaij:2012mu}. 
In this paper the same approach is used, updated to use the full 2011 data set. 

\begin{figure}[t]
\begin{center}
    \includegraphics[angle=0, width=0.45\textwidth]{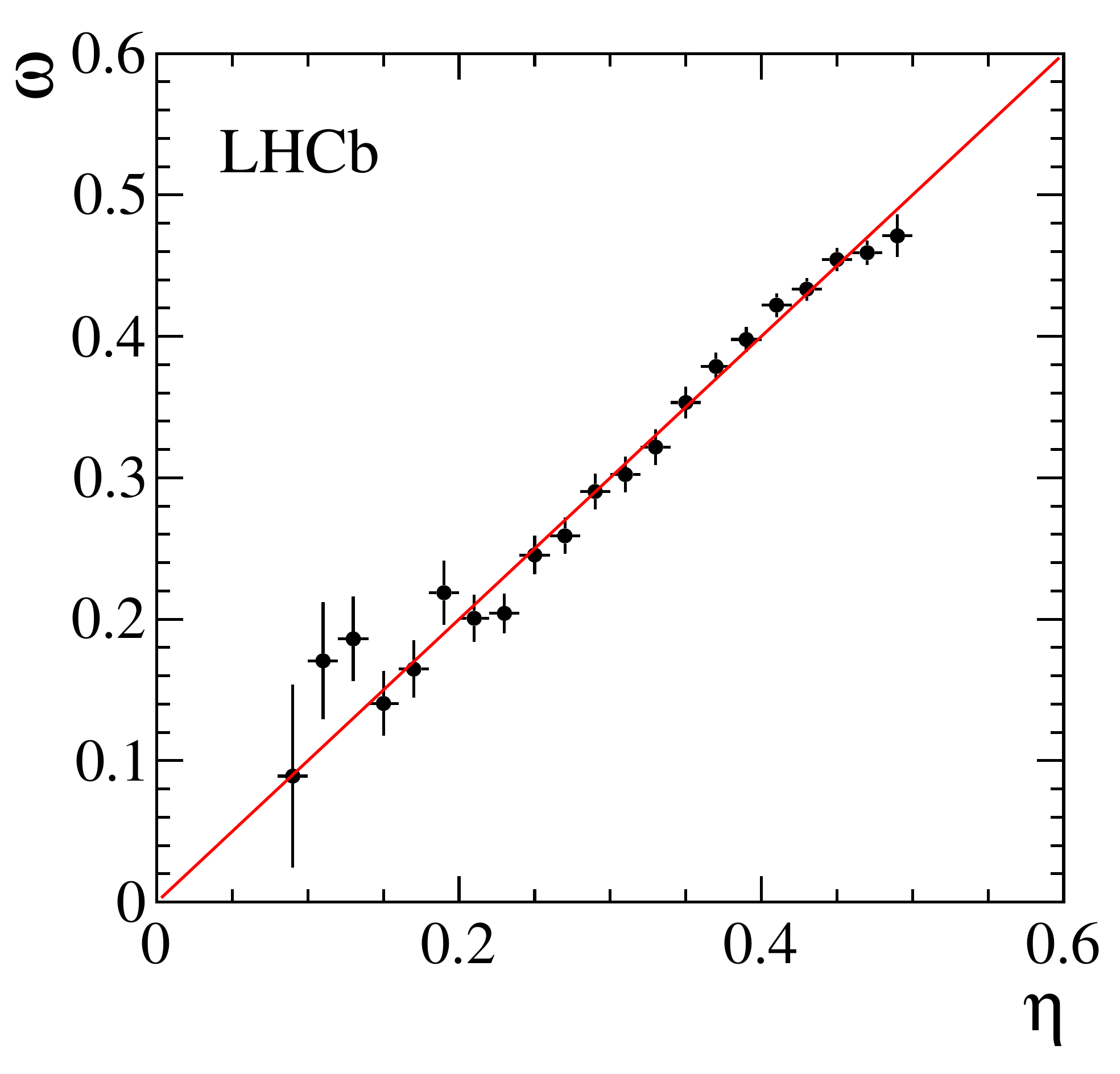}
\end{center}\vspace{-0.5cm}
\caption{\small{Average measured wrong-tag probability ($\omega$) versus estimated wrong-tag probability ($\eta$) calibrated on \BuToJPsiK signal events for the OS
tagging combinations for the background subtracted events in the signal mass window. 
Points with errors are data, the red curve represents the result of the 
wrong-tag probability calibration, corresponding to the parameters of Table~\ref{tab:Tagcal}. 
}}
\label{fig:plotCal}
\end{figure}

Calibration of the estimated wrong-tag probability, $\eta$, is performed using approximately 250\,000  $B^{+}\to\jpsi K^{+}$
 events selected from data.
The values of  $q$ and  $\eta$ measured by the OS taggers are compared to the known flavour, which
is determined by the charge of the final state kaon. 
Figure~\ref{fig:plotCal} shows the average wrong tag probability in the $B^{\pm}\to\jpsi K^{\pm}$ control channel 
in bins of $\eta$. For calibration purposes a linear relation is assumed
\begin{equation}
\label{eq:calibration}
  \begin{aligned}
\mistag(\eta) &=  & p_0 + \frac{\Delta p_0}{2} +  p_1  (\eta - \langle\eta\rangle) \,,  \\
\overline \mistag(\eta) &=  & p_0 -  \frac{\Delta p_0}{2}  + p_1  (\eta - \langle\eta\rangle) \,,
\end{aligned}
\end{equation}
where $\mistag(\eta)$ and $\overline \mistag(\eta) $ are the calibrated probabilities for wrong-tag assignment for \B and  \Bb mesons, respectively.
This parametrisation is chosen to minimise the correlation between the parameters $p_0$ and $p_1$.
The resulting values of the calibration parameters $p_0$, $p_1$, $\Delta p_0$ and $\langle\eta\rangle$ 
(the mean value of $\eta$ in the sample) are given in Table~\ref{tab:Tagcal}. The systematic uncertainties for $p_0$ and $p_1$ 
are determined by comparing the tagging performance for different decay channels, comparing different data taking periods 
and by modifying the assumptions of the fit model. 
The asymmetry parameter  $\Delta p_0$ is obtained by performing the calibration separately for $B^+$ and $B^-$ decays. 
No significant difference of the tagging efficiency or of  $p_1$ is measured ($\Delta \etag = (0.00\pm0.10)$\%, $\Delta p_1 = 0.06\pm0.04$).
Figure~\ref{fig:plotCal} shows the relation between $\omega$ and $\eta$ for the full data sample. 

The overall effective OS tagging power for $\Bs\to J/\psi K^{+}K^{-}$ candidates is  \mbox{$\effD = (2.29\pm0.06)$}\%, with 
an efficiency of $\etag=(33.00 \pm 0.28)$\% and an effective average wrong-tag probability of $(36.83\pm0.15)$\% (statistical uncertainties only).

\begin{table}[t]
\caption{\small{Calibration parameters ($p_0$, $p_1$,$\langle\eta\rangle$ and $\Delta p_0$) corresponding to the OS and SSK taggers.
The uncertainties are statistical and systematic, respectively, except for $\Delta p_0$ where they
have been added in quadrature.}}
\begin{center}
\footnotesize
\begin{tabular} { c |  c | c | c | c }
Calibration  &  $p_0$& $p_1$ &$\langle\eta\rangle$ & $\Delta p_0$\\\hline
 
OS                 & $0.392 \pm 0.002 \pm 0.008$ & $1.000 \pm 0.020 \pm 0.012$ & $0.392$  &  \phantom{+}$0.011 \pm 0.003$ \\ 
SSK  	        & $0.350 \pm 0.015 \pm 0.007$     & $1.000 \pm 0.160 \pm 0.020 $      & $0.350$  & $-0.019 \pm 0.005 $ \\ 

\end{tabular}
\end{center}

\label{tab:Tagcal}
\end{table}

\subsection{Same side kaon tagging\label{sec:SStagging}}

One of the improvements introduced in this analysis compared to Ref.~\cite{LHCb:2011aa} is the use of the SSK tagger. 
The SSK tagging algorithm was developed using large samples of simulated \Bs decays to $\Dsm\pip$ and $\jpsi\phi$
and is documented in Ref.~\cite{LHCb-CONF-2012-033}.
The algorithm preferentially selects kaons originating from the fragmentation of the signal \Bs\ meson, and rejects particles that originate 
either from the opposite-side \B decay or the underlying event. 
For the optimisation, approximately 26\,000 \BsToDsPi data events are used.
The same fit procedure employed to determine the \Bs mixing frequency \dms~\cite{Aaij:2011qx} is used to maximise the effective tagging power \effD.

The calibration was also performed using \BsToDsPi events and assuming the same linear relation 
given by Eq.~\ref{eq:calibration}.
The resulting values of the calibration parameters ($p_0,p_1, \Delta p_0$)  are given in the second row of  Table~\ref{tab:Tagcal}.
In contrast to the OS tagging case, it is more challenging to measure  $p_0$ and $p_1$ separately for true $B$ or 
$\Bb$ mesons at production using \BsToDsPi events. 
Therefore, assuming that any tagging asymmetry is caused by the difference in interaction with matter of \Kp and \Km, 
$\Delta p_0$ is estimated using $B^{+}\to\jpsi K^{-}$, where the $p$ and $\pt$ distributions of the OS tagged kaons are
first reweighted to match those of SSK tagged kaons from a large sample of fully simulated \BsToDsPi events.

The effective SSK tagging power for   $\Bs\to J/\psi K^{+}K^{-}$  events is  $\effD = (0.89\pm0.17)$\% and the tagging efficiency is $\etag=(10.26\pm0.18)$\% (statistical uncertainties only). 


\subsection{Combination of OS and SSK tagging}\label{sec:TaggingCombination}

Only a small fraction of tagged events are tagged by both the OS and the SSK algorithms.
The algorithms are uncorrelated as they select mutually exclusive charged particles, either in terms 
of the impact parameter significance with respect to the PV, or in terms of the particle identification requirements. 
The two tagging results are combined taking into account both decisions and their corresponding estimate of $\eta$. 
The combined estimated wrong-tag probability and the corresponding uncertainties are obtained by combining the individual  calibrations for the OS and SSK tagging and propagating their uncertainties according to the procedure defined in Ref.~\cite{Aaij:2012mu}. To simplify the fit implementation, the statistical and systematic uncertainties on the combined wrong-tag probability are assumed to be the same for all of these events. They are defined by the average values of the corresponding distributions computed event-by-event.
The effective  tagging power for  these OS+SSK tagged events is  $\effD = (0.51\pm0.03)$\%, and the tagging efficiency is $\etag=(3.90\pm0.11)$\%.

\subsection{Overall tagging performance}

\begin{figure}[t]
  \begin{center}
    \begin{overpic}[angle=0, width=0.45\textwidth]{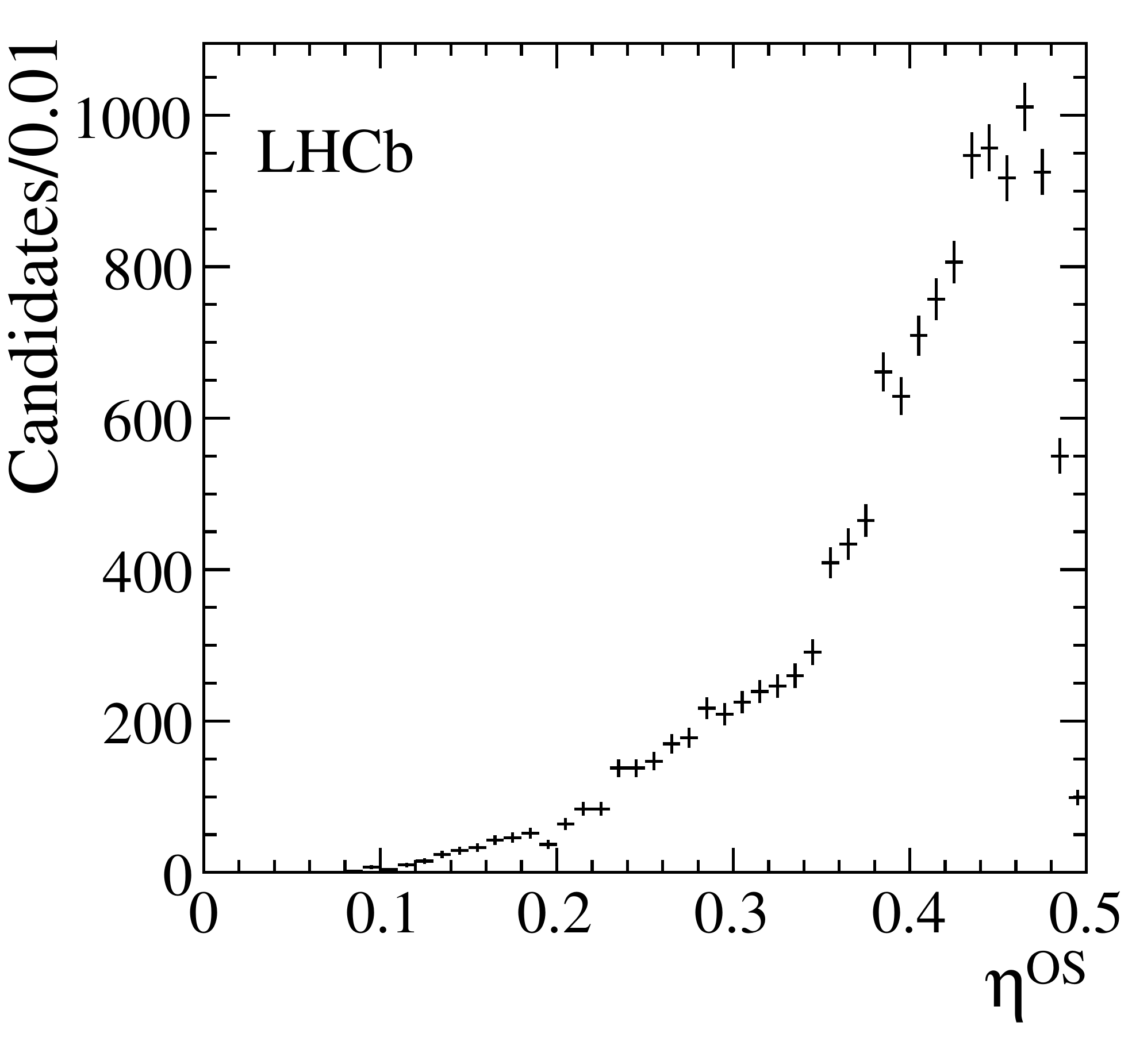}
      \put(30,55){(a)}
  \end{overpic}
    \begin{overpic}[angle=0, width=0.45\textwidth]{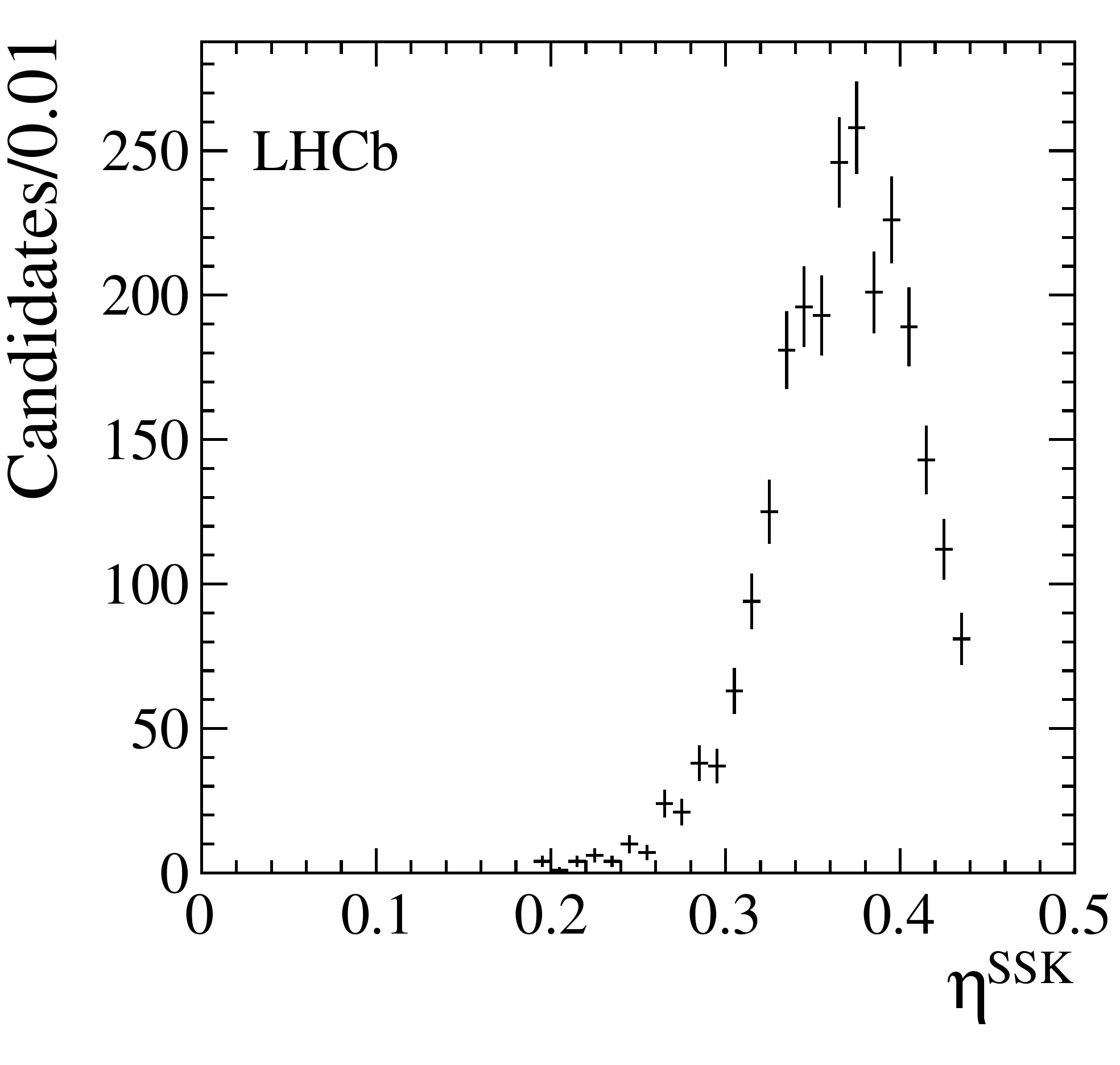}
      \put(30, 55){(b)}
  \end{overpic}
\end{center}\vspace{-0.5cm}
\caption{\small{
Distributions of the estimated wrong-tag probability, $\eta$, of the  $\Bs\to J/\psi K^{+}K^{-}$ 
signal events obtained using the \sPlot\ method 
on the $\Jpsi K^{+}K^{-}$ invariant mass distribution. Both the (a) OS-only and (b) SSK-only tagging categories are shown.}}
\label{fig:eta_jpsiphi}
\end{figure}

The overall effective  tagging power obtained by combining  all three categories   is \mbox{$\effD = (3.13 \pm 0.12 \pm 0.20)$}\%, the tagging 
efficiency is $\etag=(39.36\pm0.32)$\% and the wrong-tag probability is $\omega = 35.9$\%.
Figure~\ref{fig:eta_jpsiphi} shows the distributions of the estimated wrong-tag probability $\eta$ of the  \mbox{$\Bs\to J/\psi K^{+}K^{-}$}  signal events obtained with the \sPlot\ technique using $m(\jpsi\Kp\Km)$ as the discriminating variable. 

%
%

\clearpage

\section{Maximum likelihood fit  procedure \label{sec:fitting}}

Each event is given a signal weight, $W_i$, using the \sPlot~\cite{splot} method with $m(\jpsi\Kp\Km)$ as the discriminating variable.
A weighted fit is then performed using a signal-only PDF, denoted by $\sigpdf$, the details of which are described below. The joint negative log likelihood, $ \LLF$ constructed as 

\begin{equation}
-\ln{\cal L} =  - \alpha \sum_{\mathrm{events}\; i}{ W_i \ln{\sigpdf} },
\end{equation}
is minimised in the fit, where the factor $\alpha = \sum_i W_i  /  \sum_i W_i^2 $ is used to include the effect 
of the weights in the determination of the uncertainties~\cite{2009arXiv0905.0724X}.

\subsection{The mass model used for weighting  }

The signal mass distribution, $\sigpdf_m( m(\jpsi\Kp\Km) ;  \mBs, \sigma_m, r_{21},  f_1)$, is modelled by a double Gaussian function.
The free parameters  in the fit are the common mean, $\mBs$, the width of the narrower Gaussian function, $\sigma_m$, the ratio of the second to the first Gaussian width, $r_{21}$,  and the fraction
of the first Gaussian, $f_1$.

The background mass distribution, $\bkgpdf_m( m(\jpsi\Kp\Km) )$ is modelled by an exponential function.
The full PDF is then constructed as
\begin{equation}
\label{eqn:cFit}
\fullpdf_m= \fsig \; \sigpdf_m + ( 1 - \fsig ) \; \bkgpdf_m,
\end{equation}
where $\fsig$ is the signal fraction. Fig.~\ref{fig:mass} shows the result of fitting this model to the selected candidates.

\subsection{\boldmath Dividing the data into bins of $m(\Kp\Km)$}

The events selected for this analysis are within the  $m(\Kp\Km)$ range $[990,1050]\mevcc$.  
The data are divided into six independent sets, where the boundaries are given in Table \ref{tab:interferenceCorr}. 
Binning the data this way leads to an improvement in statistical precision by separating events with different signal fractions and 
the analysis becomes insensitive to correction factors which must be applied 
to each of the three S-wave interference terms in the differential decay rate ($f_{8}, f_{9}, f_{10}$ in Table~\ref{tab:functions}). 
These terms are required to account for an averaging effect resulting from 
the variation within each bin of the S-wave line-shape (assumed to be approximately uniform) relative to that of the P-wave (a relativistic Breit-Wigner function). In each bin, the correction factors are calculated by integrating the product of $p$ with $s^*$ which appears in 
the interference terms between the P- and S-wave, where $p$ and $s$ are the normalised $m(\Kp\Km)$ lineshapes and $^{*}$ is the complex conjugation operator,
\begin{equation}
\int_{m^L}^{m^H} { p  s^*}  \:\:{\rm d} m(\Kp\Km)
 = C_{\rm SP} e^{-i \theta_{\rm SP}},
\end{equation}
where $[m^L,m^H]$ denotes the boundaries of the $m(\Kp\Km)$ bin, 
$C_{\rm SP}$ is the correction factor and $\theta_{\rm SP}$ is absorbed in the measurements of 
$\delta_{\rm S} - \delta _{\perp}$. The $C_{\rm SP}$ correction factors are
given in Table~\ref{tab:interferenceCorr}. By using several bins these factors are close to one, whereas if 
only a single bin were used the correction would differ substantially
from one. The effect of these factors on the fit results is very small and is discussed further in
Sect.~\ref{sec:systematics}, where a different S-wave lineshape is considered.
Binning the data in $m(\Kp\Km)$ allows a repetition of the procedure described
in Ref.~\cite{LHCb-PAPER-2011-028} to resolve the ambiguous solution 
described in Sect.~\ref{sec:introduction} by inspecting the trend in the phase difference between the S- and P-wave components.

\begin{table}[t]
\caption{\small Bins of $m(\Kp\Km)$ used in the analysis and the $C_{\rm SP}$ correction
factors for the S-wave interference term, assuming a uniform distribution of non-resonant
$K^{+}K^{-}$ contribution and a non-relativistic Breit-Wigner shape for the decays via the $\phi$
resonance. \label{tab:interferenceCorr}}
\begin{center}
\begin{tabular}{c|c}
$m(\Kp\Km)$ bin [\mevcc] & $C_{\rm SP}$ \\ \hline
\phantom{0}990 -- 1008    & 0.966 \\
1008 -- 1016 & 0.956\\
1016 -- 1020 & 0.926 \\
1020 -- 1024 & 0.926\\
1024 -- 1032 & 0.956\\
1032 -- 1050 & 0.966\\

\end{tabular}

\end{center}
\end{table}

The weights, $W_i$, are determined by performing a simultaneous fit to the $m(\Jpsi K^{+}K^{-})$ distribution in each of the $m(\Kp\Km)$ bins,
using a common set of signal mass parameters and six independent background mass parameters.
This fit is performed for 
$m(\Jpsi K^{+}K^{-})$ in the range $[5200, 5550]\mevcc$ and the results for the signal mass parameters are shown in Table~\ref{tab:massparameters}.

\begin{table}[t]
\caption{\small Parameters of the common signal fit to the $m(\Jpsi K^{+}K^{-})$  distribution in data. \label{tab:massparameters}}
\begin{center}
\begin{tabular}{c|c} 
Parameter 				& Value		 \\ \hline 
$\mBs$ [\mevcc]         		&  $5368.22\phantom{0}	\pm 0.05$		\\
$\sigma_m$ [\mevcc]		&  $\phantom{000}6.08\phantom{0}		\pm 0.13$		\\
$f_1$ 	     				&  $\phantom{0000}0.760		\pm 0.035$		\\
$r_{21}$ 					&  $\phantom{000}2.07\phantom{0}		\pm 0.09$		\\

\end{tabular}

\end{center}
\end{table}

\subsection{The signal PDF}

The physics parameters of interest in this analysis are  
$\Gs$,  $\DGs$,  $\azerosq$, $\aperpsq$, $\fS$, $\delpar$, $\delperp$, $\dels$, $\phis$, $\maglambda$ and $\dms$,
all of which are defined in Sect.~\ref{sec:pheno}.
The signal PDF, $\sigpdf$,  is a function of the decay time, $t$, and angles, $\Omega$, 
and is conditional upon the estimated wrong-tag probability for the event, $\tagomega$, and the estimate of the decay time resolution for the event, $\eventres$. 
The data are separated into disjoint sets corresponding to each of the possible tagging decisions $q\in\{-1,0,+1\}$ and the unbiased and biased trigger samples. A separate signal PDF, 
$\sigpdf_q(t, \Omega  | \eventres, \tagomega ; Z,N )$, is constructed for each event set, where 
$Z$ represents the physics parameters and $N$ represents nuisance parameters described above. 

The $\sigpdf_q$  are constructed from the differential decay rates of $\Bs$ and $\Bsb$ mesons described in Sect. \ref{sec:pheno}. 
Denoting 
$\frac{\deriv^{4} \Gamma(\Bs \to \jpsi KK) }{\deriv t \;\deriv\Omega} $ 
by $X$ and $\frac{\deriv^{4} \Gamma(\Bsb \to \jpsi KK) }{\deriv t \;\deriv\Omega} $ by $\overline X$, then
\begin{equation}
\sigpdf_q    =    \frac{ s_q} { \int s_q \: \deriv t \;\deriv\Omega},  \\
\end{equation}
where
\begin{eqnarray}\nonumber
s_{+1}  &=   &   \Big[ [ \:(1 - \wtrue) \; X( t, \Omega ; Z)+  \wtruebar \; \overline X( t, \Omega ; Z) \:  ]  \otimes  R(t; \eventres) \Big] \; \varepsilon_t(t)  \; \varepsilon_\Omega(\Omega),   \\
s_{-1}  &=   &  \Big[ [\:  \wtrue \; X( t, \Omega ; Z)  +  (1- \wtruebar)  \; \overline X( t, \Omega ; Z) \: ] \otimes  R(t; \eventres)  \Big]   \; \varepsilon_t(t) \; \varepsilon_\Omega(\Omega),  \\ 
s_{0}  &=   &   \frac{1}{2} \Big[  [\: X( t, \Omega ; Z)  +  \overline X( t, \Omega ; Z) \: ]   \otimes  R(t; \eventres)  \Big] \; \varepsilon_t(t) \; \varepsilon_\Omega(\Omega). \ \nonumber
\end{eqnarray}
Asymmetries in the tagging efficiencies and relative magnitudes of the production rates for $\Bs$ and $\Bsb$ mesons, as well as  
the factor $|p/q|^2$  are not included in the model. 
Sensitivity to these effects is reduced by the use of separately normalised PDFs for each of the tagging decisions and any residual effect is shown to be negligible.

All physics parameters are free in the fit apart from $\dms$, which is constrained to the value measured by  \lhcb of
$17.63\pm 0.11 \invps$~\cite{Aaij:2011qx}. The parameter $\delsperp$ is used in the minimisation instead of
$\dels$ as there is a large (90\%) correlation between $\dels$ and $\delperp$.

In these expressions the terms $\wtrue$ and $\overline \omega$
represent the wrong-tag probabilities for a candidate produced as a genuine $\Bs$ or $\Bsb$ meson, respectively, and are a function of  $\tagomega$ and the (nuisance) calibration
parameters $(p_1, p_0, \langle\eta\rangle, \Delta p_0)$  as given in Eq.~\ref{eq:calibration}. The calibration parameters are given in Table~\ref{tab:Tagcal} and are all included in the fit via Gaussian constraints with widths equal to their uncertainties. 

The expressions are  convolved with the decay time resolution function, $R(t; \eventres)$ (Sect.~\ref{sec:ptresolution}).
The scale factor parameter, $\eventresscale$, is included in the fit with its value constrained by a Gaussian constraint with width equal to its uncertainty. The $\varepsilon_t(t)$  and $\varepsilon_\Omega(\Omega)$ terms are the decay time acceptance and decay-angle acceptance, respectively. The two different trigger samples have different decay time acceptance functions. These are described in Sect.~\ref{sec:acceptance}.

Since this weighted fit uses only a signal PDF there is no need to include the distributions of either the estimated wrong tag probability,  
$\tagomega$, or the decay time resolution for each event, $\eventres$. 
The physics parameter estimation is then performed by a simultaneous fit to  the weighted data in each of the $m(\Kp\Km)$ bins
for each of the two trigger samples.
All parameters are common, except for the S-wave fraction $F_{\rm S}$ and the phase difference 
$\delta_{\rm S}-\delta_{\perp}$, which are independent parameters for each range.

%
%

\section{\boldmath Results for $\BtoJpsiKK$ decays \label{sec:results}}

The results of the fit for the principal physics parameters are given in Table~\ref{tab:results} 
for the solution with $\DGs>0$,
showing both the statistical and the total systematic uncertainties described in Sect.~\ref{sec:systematics}.

\begin{figure}

\centering
		\includegraphics[trim=12mm 11mm 16mm 23mm, clip=true, width=0.495\textwidth]{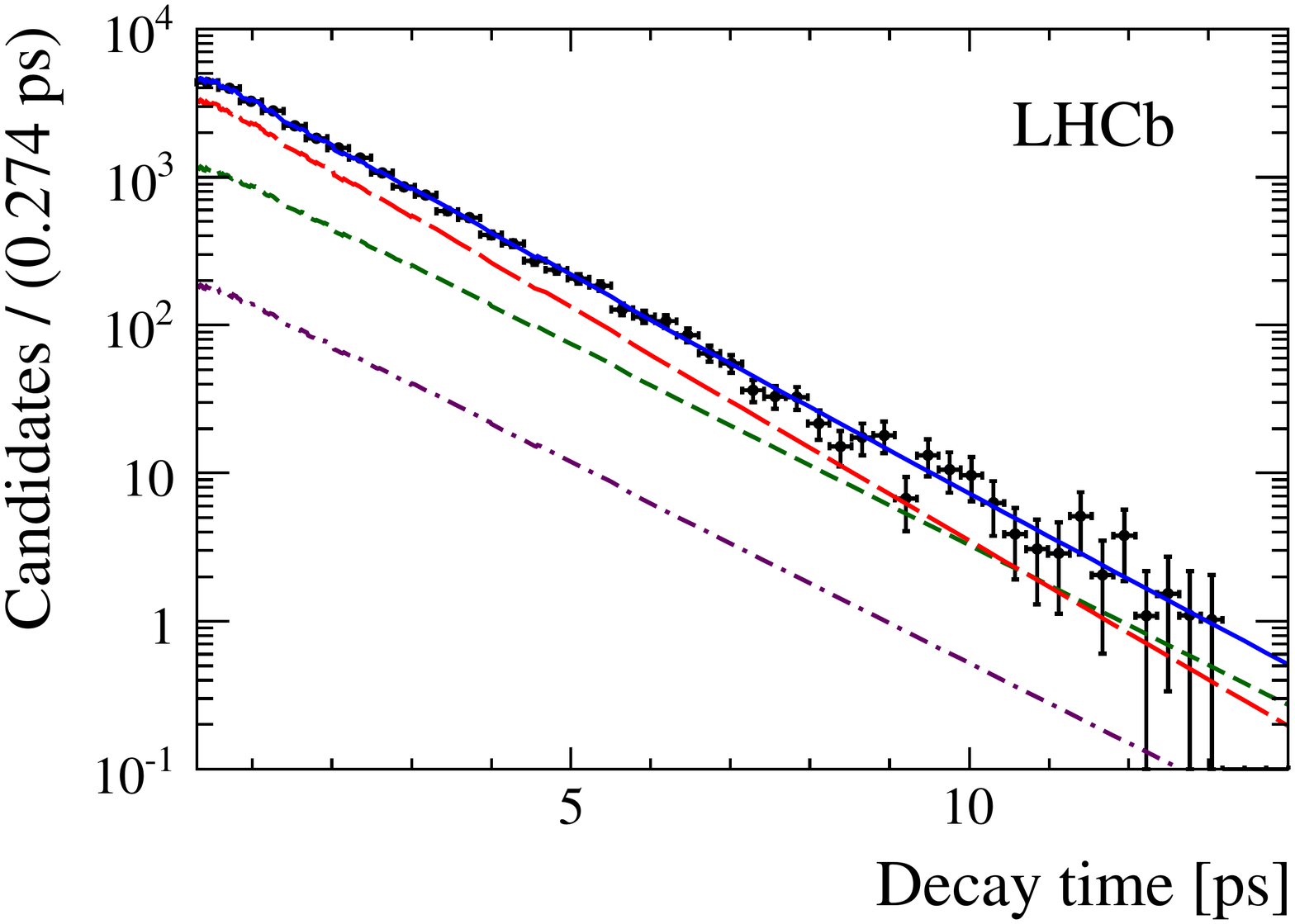}%
                \hspace*{0.01\textwidth}%
		\includegraphics[trim=12mm 11mm 16mm 23mm, clip=true, width=0.495\textwidth]{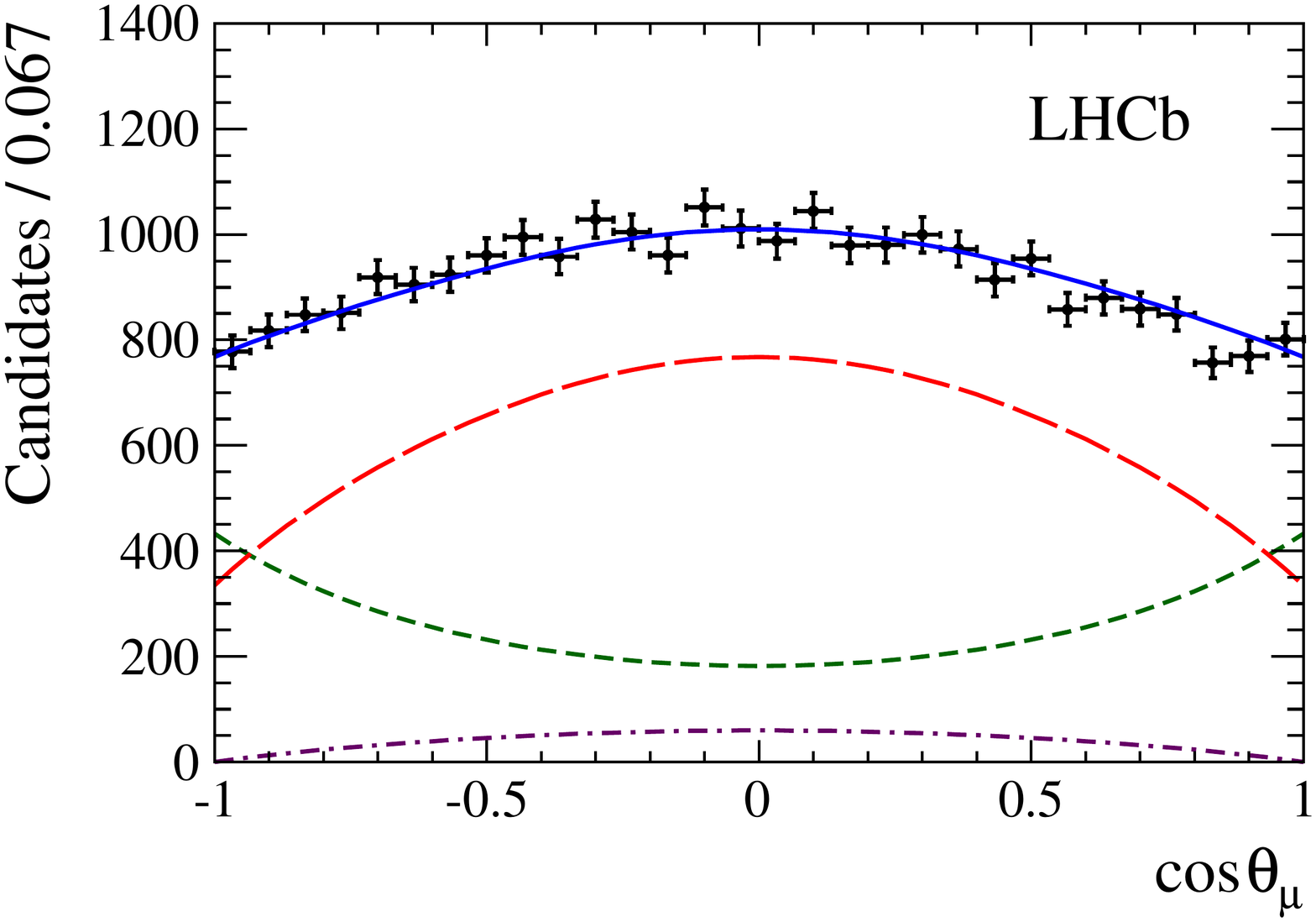}
		\includegraphics[trim=12mm 11mm 16mm 23mm, clip=true, width=0.495\textwidth]{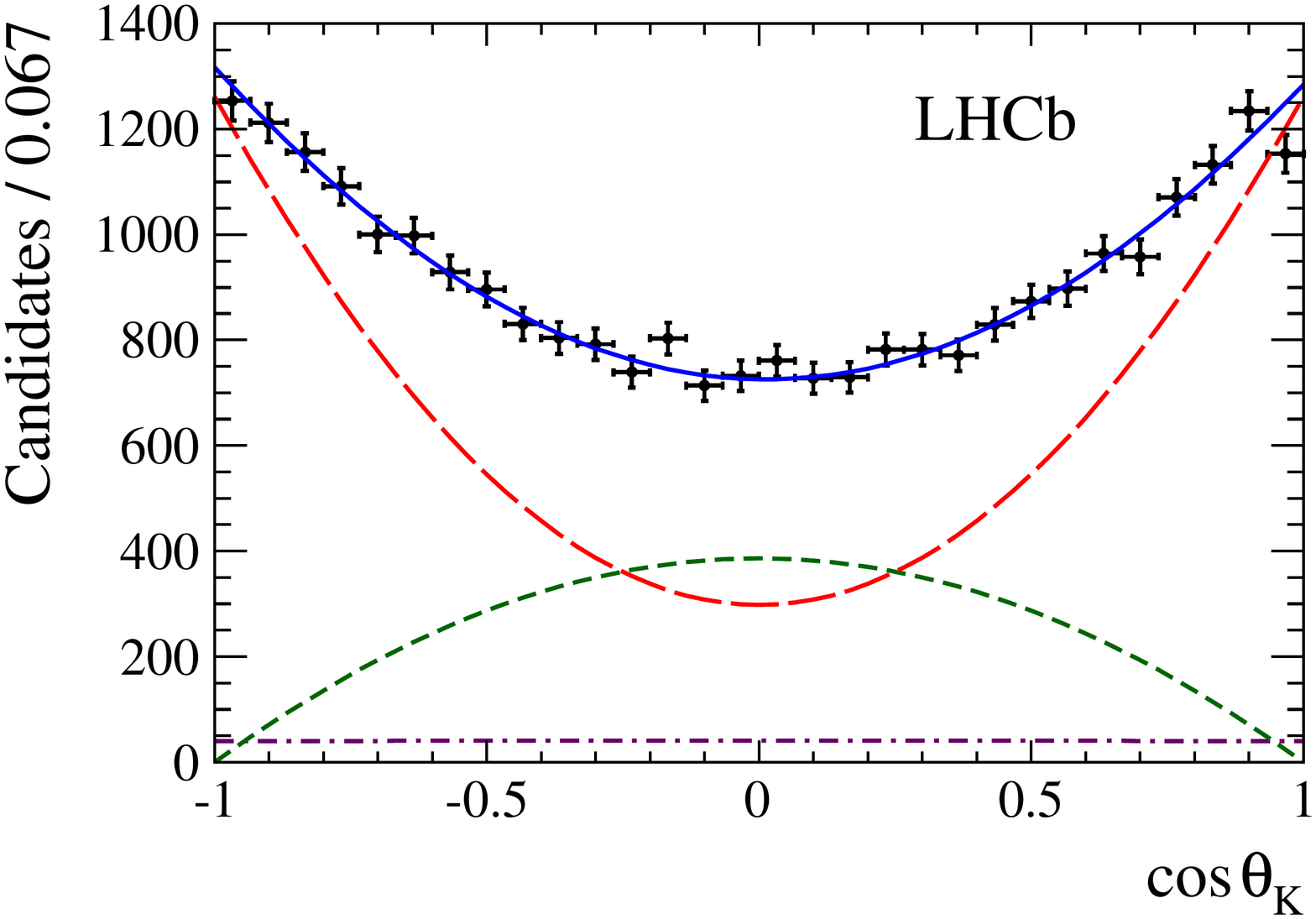}%
                \hspace*{0.01\textwidth}%
 		\includegraphics[trim=12mm 11mm 16mm 23mm, clip=true, width=0.495\textwidth]{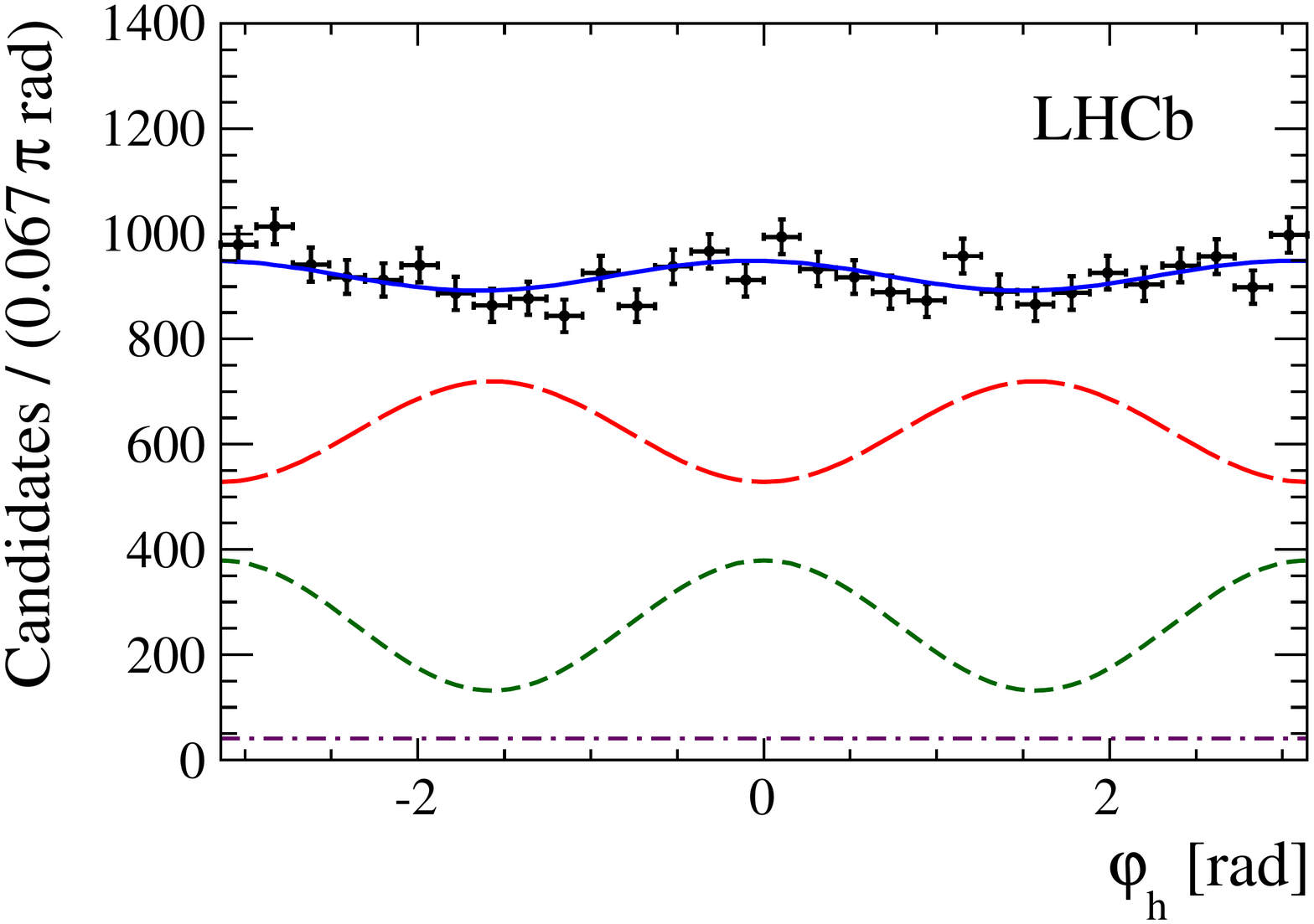}
\caption{\small\label{fig:results-projections} 
Decay-time and helicity-angle distributions for $\Bs\to\jpsi\Kp\Km$ decays (data points) with the one-dimensional projections of the PDF at the maximal likelihood point. 
The solid blue line shows the total signal contribution, which is composed of 
\CP-even (long-dashed red), \CP-odd (short-dashed green) and S-wave (dotted-dashed purple) contributions.}
\end{figure}

The statistical correlation matrix is shown in Table~\ref{tab:results-corr}. 
The projections of the decay time and angular distributions are shown in Fig.~\ref{fig:results-projections}.
It was verified that the observed uncertainties are compatible with the expected sensitivities,  by generating and fitting to a large number of
simulated experiments.

\begin{figure}
	\begin{center}
          \includegraphics[scale=0.62]{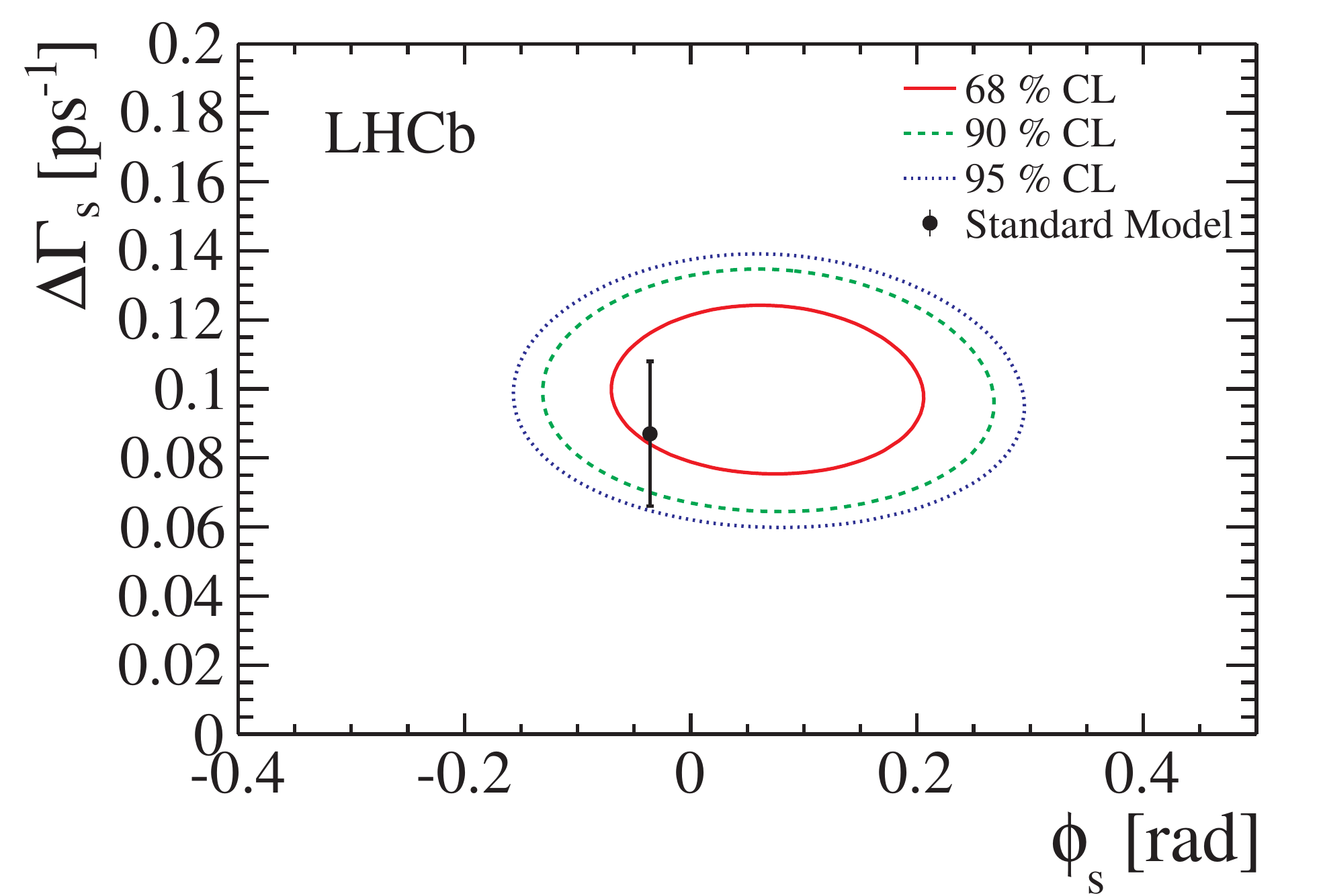}
       \end{center}
	\caption{\small\label{fig:2Dscan-1} Two-dimensional profile likelihood in the ($\DGs$, $\phis$) plane for
	the \mbox{$\Bs\to\jpsi\Kp\Km$} dataset. Only the statistical uncertainty is included.
	The SM expectation of \mbox{$\DGs=0.087 \pm 0.021\invps$}
	and $\phis = -0.036 \pm 0.002\rad$ is shown as the black point with error bar~\cite{CKMfitter,Lenz:2006hd}.}
\end{figure}

Figure~\ref{fig:2Dscan-1} shows the 68\%, 90\% and 95\% CL contours obtained from the  two-dimensional profile likelihood ratio in the
($\DGs$, $\phis$) plane, corresponding to decreases in the log-likelihood of 1.15, 2.30 and 3.00 respectively.
Only statistical uncertainties are included.
The SM expectation~\cite{Lenz:2006hd,*Badin:2007bv,*Lenz:2011ti} is shown.

\begin{figure}
        \centering
        \includegraphics[trim=22mm 13mm 16mm 29mm, clip=true, width=0.7\textwidth]{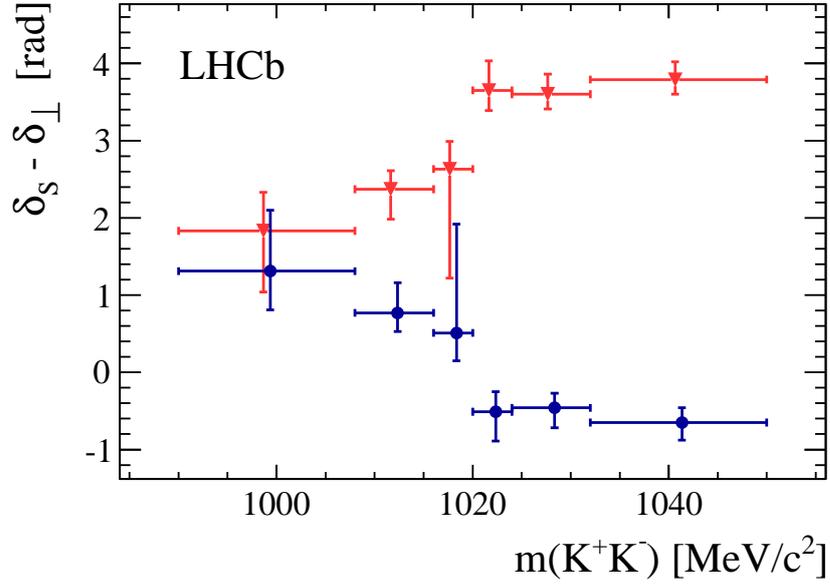}
	\caption{\small\label{fig:swavephase}  
	Variation of $\dels-\delperp$ with $m(\Kp\Km)$ where the uncertainties are the quadrature sum of the statistical and systematic uncertainties in each bin. 
	The decreasing phase trend (blue circles) corresponds to the physical solution with \phis\ close to zero and $\Delta\Gamma_{s}>0$. 
	The ambiguous solution is also shown.             }
\end{figure}

The results for the S-wave parameters are shown in Table~\ref{tab:results-phase}. 
The likelihood profiles for these parameters are non-parabolic and are asymmetric.
Therefore the 68\% CL intervals obtained from the likelihood profiles, corresponding 
to a decrease of 0.5 in the log-likelihood, are reported.
The variation of $\dels - \delperp$ with $m(\Kp\Km)$ is shown in Fig.~\ref{fig:swavephase}. 
The decreasing trend confirms that expected for the physical solution with $\phis$ close to zero, as found in Ref.~\cite{LHCb-PAPER-2011-028}.

All results have been checked by splitting the dataset into sub-samples to compare different data taking periods, 
magnet polarities, \Bs-tags and trigger categories. In all cases the results are consistent between the independent sub-samples.
The measurements of $\phis$, $\DGs$ and $\Gs$ are the most precise to date. 
Both $\DGs$ and $\phis$ agree well with the SM expectation~\cite{CKMfitter,Lenz:2006hd}.

\begin{figure}
	\begin{center}
          \includegraphics[scale=0.6]{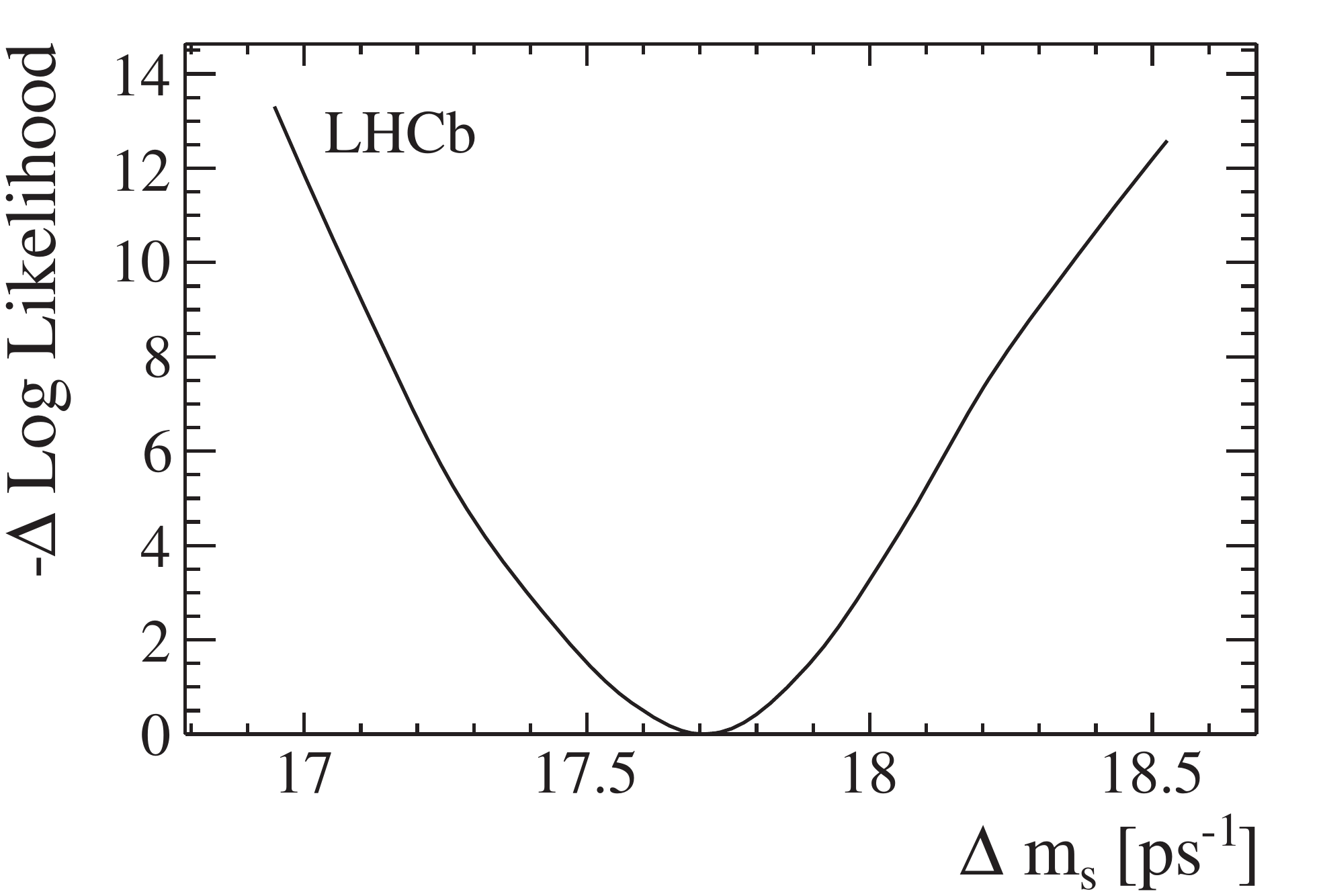}
       \end{center}
	\caption{\small\label{fig:dmsplot}   Profile likelihood for $\dms$  from a fit where $\dms$ is unconstrained.}
\end{figure}

These data also allow an independent measurement of \dms without constraining it to the value reported in Ref.~\cite{Aaij:2011qx}.
This is possible because there are several terms in the differential decay rate of Eq.~\ref{eq:Eqbsrate}, principally $h_4$ and $h_6$, which contain sinusoidal terms in
$\dms t$ that are not multiplied by $\sin\phis$. 
Figure~\ref{fig:dmsplot} shows the likelihood profile as a function of $\dms$ from a fit to the data where $\dms$ is not constrained. The result 
of the fit gives
\begin{displaymath}
\dms = 17.70   \pm   0.10 \;\text{(stat)}   \pm 0.01 \;\text{(syst)\invps},
\end{displaymath}
\noindent which is consistent with other measurements~\cite{Aaij:2011qx,Abazov:2006dm,Abulencia:2006mq,LHCb-PAPER-2013-006}.

\begin{table}[t]
\caption{\small Results of the maximum likelihood fit for the principal physics parameters. The first uncertainty is statistical and the second is systematic.
The value of $\dms$ was constrained to the measurement reported in Ref.~\cite{Aaij:2011qx}.
The evaluation of the systematic uncertainties is described in Sect.~\ref{sec:systematics}.}
\centerline{
    \begin{tabular}{l | c}
                                Parameter    &     Value\\
      \hline
                                $\Gs$ [\invps]    	&   $0.663   \pm  0.005        \pm  0.006$  \\ 
                                $\DGs$ [\invps]   	&   $0.100    \pm  0.016        \pm  0.003$   \\
                                $\aperpsq$       	&   $0.249   \pm  0.009         \pm  0.006$  \\
                                $\azerosq$        	&   $0.521   \pm  0.006          \pm  0.010$  \\
                                $\delpar$~[rad]   	&   $3.30\,     ^{+ 0.13}_{- 0.21}   \pm  0.08$    \\
                                $\delperp$~[rad]  	&   $3.07     \pm  0.22              \pm     0.08$    \\
                                $\phis$~[rad]      	&   $0.07    \pm  0.09               \pm  0.01$   \\
                                $\maglambda$    	&   $0.94    \pm  0.03                \pm  0.02$   \\ 
    \end{tabular}
     }
\label{tab:results}
\end{table}

\begin{table}[t]
\caption{\small Correlation matrix for the principal physics parameters.}
\begin{center}\small
\begin{tabular}{l|c|c|c|c|c|c|c|c} 
                & $\Gs$ &  $\DGs$   & $\aperpsq$   & $\azerosq$   &  $\delpar$    &  $\delperp$  &  $\phis$    &  $\maglambda$  \\
                & [\invps] &  [\invps]   &    &    &  [rad]    &  [rad]  &  [rad]     &    \\
  \hline
  $\Gs$ [\invps]       &  $\phantom{+}1.00$  &  ${ -0.39}$  &  $\phantom{+}{ 0.37}$  & ${ -0.27}$  &  $-0.09$        &  $-0.03$       &  $\phantom{+}0.06$        &  $\phantom{+}0.03$          \\
  $\DGs$ [\invps]        &        &  $\phantom{+}1.00$        &  ${-0.68}$  &  $\phantom{+}{ 0.63}$  &  $\phantom{+}0.03$         &  $\phantom{+}0.04$        &  $-0.04$       &  $\phantom{+}0.00$          \\
  $\aperpsq$    &        &              &  $\phantom{+}1.00$        & ${ -0.58}$  &  ${ -0.28}$  &  $-0.09$       &  $\phantom{+}0.08$        &  $-0.04$         \\
  $\azerosq$    &        &              &              &  $\phantom{+}1.00$        &  $-0.02$        &  $-0.00$       &  $-0.05$       &  $\phantom{+}0.02$         \\
  $\delpar$~[rad]     &        &              &              &              &  $\phantom{+}1.00$        &  $\phantom{+}{ 0.32}$  &  $-0.03$       &  $\phantom{+}0.05$          \\
  $\delperp$~[rad]    &        &              &              &              &               &   $\phantom{+}1.00$       &  $\phantom{+}{ 0.28}$  &  $\phantom{+}0.00$          \\
  $\phis$~[rad]       &        &              &              &              &               &              &  $\phantom{+}1.00$        &  $\phantom{+}0.04$          \\
  $\maglambda$  &        &              &              &              &               &              &              &  $\phantom{+}1.00$          \\
 \end{tabular}
\end{center}
\label{tab:results-corr}
\end{table}

\begin{table}
\caption{\small Results of the maximum likelihood fit for the S-wave parameters, with asymmetric statistical and symmetric systematic uncertainties.
The evaluation of the systematic uncertainties is described in Sect.~\ref{sec:systematics}. }
\begin{center}
\begin{tabular}{c|c|c|c|c}
$m(\Kp\Km)$ bin [\mevcc]     &Parameter                       	& Value 	& $\sigma_\text{stat}$ (asymmetric)   & $\sigma_\text{syst}$   \\ 
\hline
$\phantom{0}990-1008$ &	$\fS$                    				& 0.227 	& ${+0.081} , {-0.073}$	&   0.020 \\
                   &	$\delsperp$     [rad]            			& 1.31 	& $+0.78, -0.49$		&    0.09\\ 
\hline
$1008-1016$ &	$\fS$                    			& 0.067 	& $+0.030, -0.027$ 		&    0.009 \\
                     &	$\delsperp$     [rad]            			& 0.77 	& $+0.38, -0.23$ 		&    0.08 \\  
\hline
$1016-1020$ &	$\fS$                    			& 0.008 	& $+0.014, -0.007$ 		&    0.005\\
                    &	$\delsperp$    [rad]            			& 0.51 	& $+1.40, -0.30$ 		&    0.20\\  
\hline
$1020-1024$ &	$\fS$                    			& 0.016 	& $+0.012, -0.009$ 		&    0.006\\
                     &	$\delsperp$     [rad]            			& $-0.51$ 	& $+0.21, -0.35$ 		&    0.15\\  
\hline
$1024-1032$ &	$\fS$                    			& 0.055 	& $+0.027, -0.025$ 		&    0.008\\
                      &	$\delsperp$     [rad]            			& $-0.46$ 	& $+0.18, -0.26$ 		&    0.05\\ 
\hline
$1032-1050$ &	$\fS$                    			& 0.167 	& $+0.043, -0.042$ 		&   0.021 \\ 
                     &	$\delsperp$     [rad]            			& $-0.65$ 	& $+0.18, -0.22$ 		&    0.06 \\	 
\end{tabular}
\end{center}

\label{tab:results-phase}
\end{table}

\clearpage

%
%

\section{\boldmath Systematic uncertainties for \mbox{$\BtoJpsiKK$} decays\label{sec:systematics}}

The parameters $\dms$,  the tagging
calibration parameters, and the event-by-event proper time scaling factor, $\eventresscale$,
are all allowed to vary within their uncertainties in the fit. 
Therefore the systematic uncertainties from these sources are included in the statistical uncertainty on the physics parameters.
The remaining systematic effects are discussed below and summarised in Tables~\ref{tab:finalsystematicssummary}, \ref{tab:finalsystematicssummary_FS} and  \ref{tab:finalsystematicssummary_deltaS}.

The parameters of the $m(\Jpsi K^{+}K^{-})$ fit model are varied within their uncertainties and a new set of event weights are calculated. Repeating the full decay time and angular fit using the new weights gives negligible differences with respect to the results of the nominal fit. The assumption that $m(\Jpsi K^{+}K^{-})$ is independent of the decay time and angle variables is tested by re-evaluating the weights in bins of the decay time and angles. Repeating the full fit with the modified weights gives new estimates of the physics parameter values in each bin. The total systematic uncertainty is computed from the square root of the sum of the individual variances, weighted by the number of signal events in each bin in cases where a significant difference is observed.

Using simulated events, the only identified peaking background is from \mbox{$\Bd\to\jpsi K^{*}(892)^{0}$} events where the
pion from the $K^{*}(892)^{0}$ decay is misidentified as a kaon. 
The fraction of this contribution was estimated from the simulation to be at most
1.5{\%} for $m(\Jpsi K^{+}K^{-})$ in the range $[5200,5550]\mevcc$.
The effect of this background (which is not included in the PDF modelling) was estimated by embedding the simulated
$\Bd\to\jpsi K^{*}(892)^{0}$ events in the signal sample and repeating the fit. The resulting variations are taken as systematic uncertainties.
The contribution of \Bs mesons coming from the decay of $B_{c}^{+}$ mesons is estimated to be negligible.

\begin{figure}[tb]
  \begin{center}
    \begin{overpic}[width=0.41\textwidth]{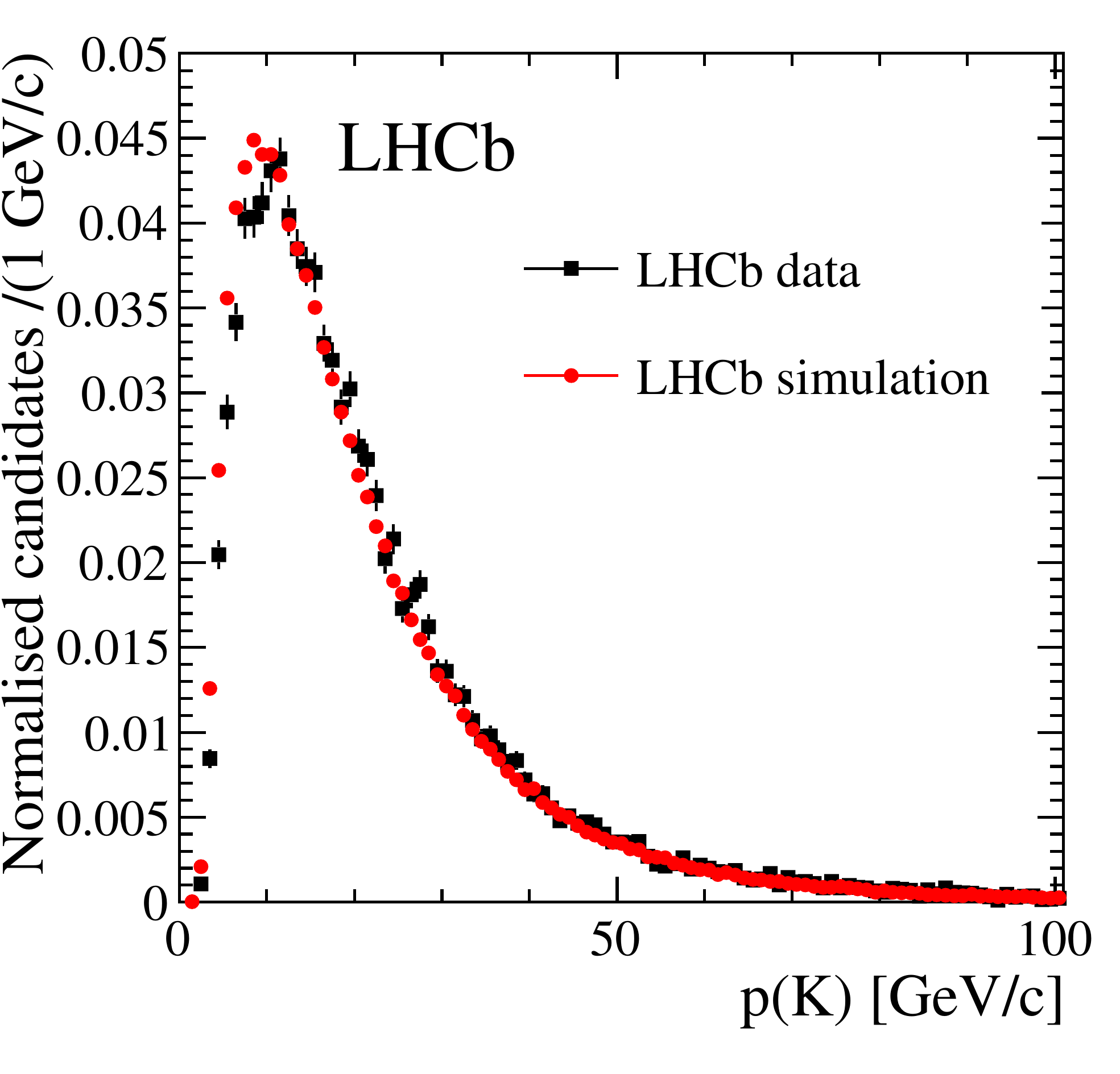}
          \put(64,41){(a)}
  \end{overpic}
    \begin{overpic}[width=0.41\textwidth]{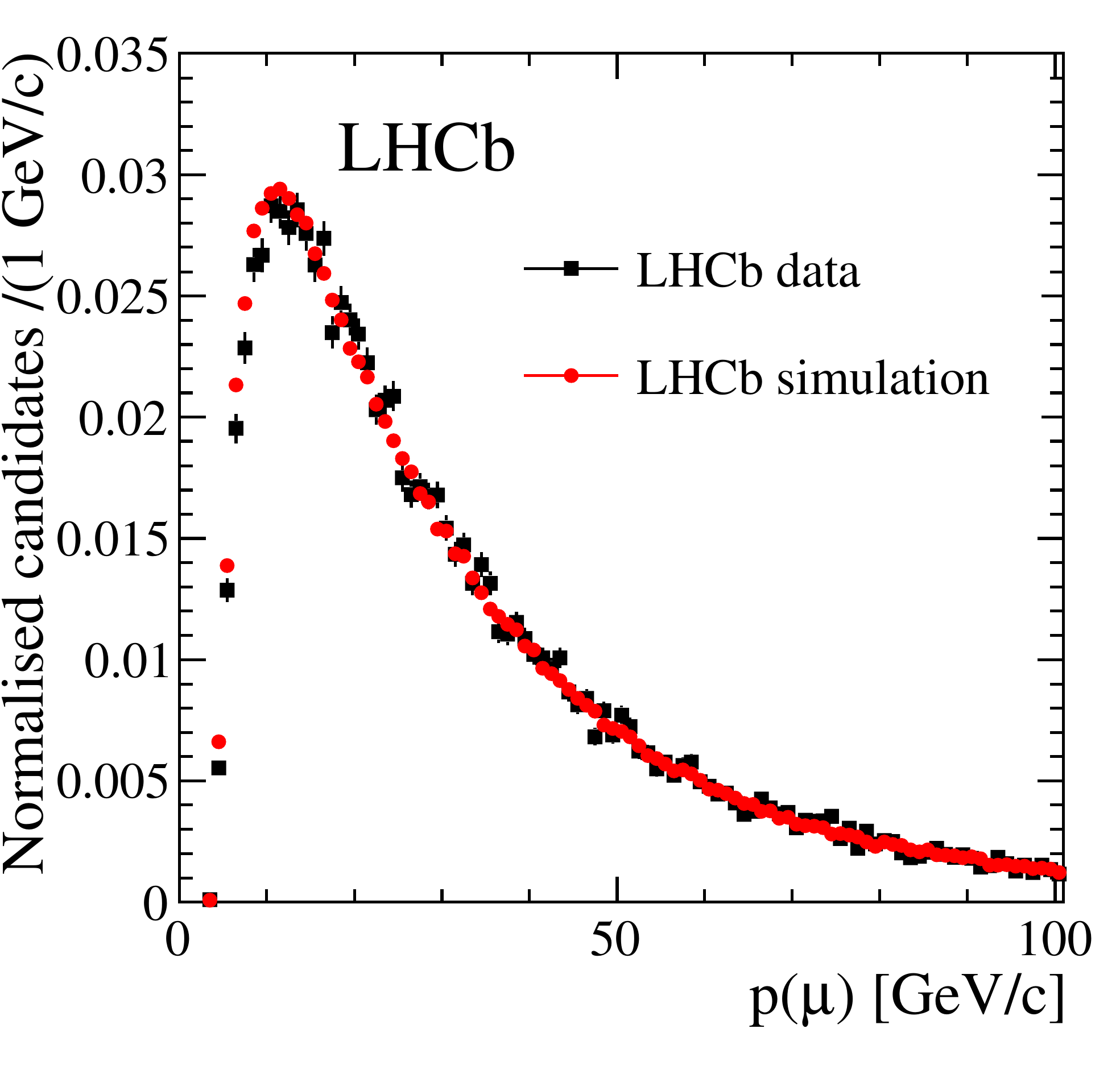}
          \put(64,41){(b)}
  \end{overpic}
  \end{center}\vspace{-0.5cm}
  \caption{\small Background-subtracted (a) kaon and (b) muon momentum distributions for \mbox{$B_{s}^{0}\to\jpsi\Kp\Km$} signal events in data
    compared to simulated $B_{s}^{0}\to\jpsi\phi$ signal events. The distributions are normalised to the same area. 
    A larger deviation is visible for kaons.}
  \label{fig:AcceptKaonMom}
\end{figure}

Since the angular acceptance function, $\effa$, is determined from simulated events, it is important that the simulation gives a good description of  the dependence of final-state particle efficiencies on their kinematic properties. Figure~\ref{fig:AcceptKaonMom} shows significant discrepancies
between simulated $B_{s}^{0}\to\jpsi\phi$ events and selected $B_{s}^{0}\to\jpsi\Kp\Km$ data events where the background has been
subtracted. 
To account for this difference the simulated events are re-weighted such that the 
kaon momentum distribution matches the data (re-weighting the muon momentum has negligible effect).
A systematic uncertainty is estimated by determining $\effa$ after this re-weighting
and repeating the fit. The changes observed in physics parameters are taken as  systematic uncertainties.
A systematic uncertainty is included which arises from the limited size of the simulated data sample used to determine $\effa$.

The lower decay time acceptance is included in the PDF using the binned functions described in Sect.~\ref{sec:acceptance}. A systematic uncertainty is determined by repeating the fits with the bin values varied randomly within their statistical precision.
The standard deviation of the distribution of central values obtained for each fit parameter is then assigned as the systematic uncertainty.
The slope  of the acceptance correction at large lifetimes is $\beta=(-8.3 \pm 4.0)\times10^{-3}\invps$. This 
leads to a $4.0\times 10^{-3}\invps$ systematic uncertainty \mbox{on $\Gamma_{s}$.}

The uncertainty on the LHCb length scale is estimated to be at most 0.020\%,
which translates directly in an uncertainty on \Gs{} and \DGs{} of 0.020\%
with other parameters being unaffected.
The momentum scale uncertainty is at most 0.022\%. As it affects both the reconstructed 
 momentum and mass of the \Bs meson, it cancels to a large extent and the resulting
effect on \Gs{} and \DGs{} is negligible.

The $C_{\rm SP}$ factors (Table~\ref{tab:interferenceCorr}) used in the nominal fit assume a non-resonant shape for the S-wave contribution. As a cross-check the factors are re-evaluated assuming a Flatt\'{e} shape~\cite{flatte} and the fit is repeated. There is a negligible effect on all physics parameters except $\delta_{\rm S}-\delta_{\perp}$.
A small shift (approximately 10\% of the statistical uncertainty) is observed in $\delta_{\rm S}-\delta_{\perp}$ in each bin of $m(\Kp\Km)$, 
and is assigned as a systematic uncertainty.

A possible bias of the fitting procedure is investigated by generating and fitting many simplified simulated experiments
of equivalent size to the data sample. The resulting biases are small, and those which are not compatible with zero within three standard deviations are quoted as systematic uncertainties.

The small offset, $d$, in the decay time resolution model was set to zero during the fitting procedure. A 
corresponding systematic uncertainty was evaluated using simulated experiments and found to be negligible for 
all parameters apart from $\phi_s$ and $\delperp$.

A measurement of the asymmetry that results from \CP{} violation in the
interference between \Bs--\Bsb{} mixing and decay is potentially
affected by \CP{} violation in the mixing, direct \CP violation in the decay, production asymmetry and tagging asymmetry. 
In the previous analysis~\cite{LHCb:2011aa} an explicit systematic uncertainty was included to account for this.
In this analysis the fit parameter $\maglambda$ is added, separate tagging calibrations are used for $\Bs$ and \Bsb decisions, 
as well as separate normalisations of the PDF for each tagging decision. Any residual effects due to  tagging 
efficiency asymmetry and production asymmetry are shown to be negligible through
simulation studies.

The measurement of $\dms$ determined from these data alone without applying a constraint has been reported in
Sect.~\ref{sec:results}.
The dominant sources of systematic uncertainty come from the knowledge of the LHCb length and momentum scales. 
No significant systematic effect is observed after varying the decay time and angular acceptances and the
decay time resolution. Adding all contributions in quadrature gives a  total systematic uncertainty of $\pm0.01\invps$.

{
\begin{table}[tb]
  \caption{\small Statistical and systematic uncertainties.}
  \begin{center}\footnotesize
    \begin{tabular}{l|c|c|c|c|c|c|c|c}
      Source            						& $\Gs$ 		& $\DGs$   	& $\aperpsq$ 	& $\azerosq$ 	& $\delpar$ 	&$\delperp$	& $\phi_s$  	& $\maglambda$   	\\
       									& [ps$^{-1}$]	&[ps$^{-1}$]	&  			& 			&  [rad] 		&  [rad] 		& [rad] 		&				\\[0.5ex] \hline
      \rule[3.5mm]{0mm}{0mm}Stat. uncertainty                       			& 0.0048 		& 0.016 		& 0.0086 		& 0.0061 		& $^{+0.13}_{-0.21}$ & 0.22 	& 0.091 		& 0.031 			\\[0.1ex]
       \hline
      Background subtraction			    	& 0.0041 		& 0.002		& --			& 0.0031		& 0.03 		& 0.02 		& 0.003		& 0.003  			\\
      \BdToJPsiKst{} background     			& --			& 0.001 		& 0.0030		& 0.0001 		& 0.01 		& 0.02 		& 0.004 		& 0.005 			\\ 
      Ang. acc. reweighting 					& 0.0007 		& -- 			& 0.0052		& 0.0091		&0.07 		& 0.05		& 0.003		& 0.020			\\ 
      Ang. acc. statistical  					& 0.0002 		& --			& 0.0020		& 0.0010		& 0.03		& 0.04		& 0.007		& 0.006			\\ 
      Lower decay time acc. model            		& 0.0023 		& 0.002 		& -- 			& -- 			& -- 			& -- 			& -- 			& --  				\\ 
      Upper decay time acc. model 			& 0.0040		& --			&-- 			&-- 			&-- 			&-- 			&-- 			&--  				\\   
      Length and mom. scales               		& 0.0002 		& -- 			& -- 			& -- 			& --	 		& -- 			& -- 			& -- 				\\ 
      Fit bias 							& -- 			& -- 			& 0.0010 		& -- 			& -- 			&-- 			& --			& -- 				\\ 
      Decay time resolution offset				& -- 			& -- 			& --	 		& -- 			& -- 			& 0.04		& 0.006		& -- 				\\ \hline
      Quadratic sum of syst.     				& 0.0063		& 0.003		& 0.0064		& 0.0097		& 0.08		& 0.08		& 0.011		& 0.022 			\\ 
      Total uncertainties     					& 0.0079		& 0.016		& 0.0107		& 0.0114		& $^{+0.15}_{-0.23}$& 0.23	& 0.092		& 0.038			\\ 
      \end{tabular}
  \end{center}
  \label{tab:finalsystematicssummary}

\end{table}
}

\begin{table}[tb]
  \caption{\small Statistical and systematic uncertainties for S-wave fractions in bins of $m(K^{+}K^{-})$.}
  \begin{center}\footnotesize
    \begin{tabular}{l|c|c|c|c|c|c}
          Source                                        &  bin 1   &  bin 2   & bin 3   &  bin 4   &  bin 5   &  bin 6   \\ 
                                              & $F_{\rm S}$           & $F_{\rm S}$           & $F_{\rm S}$          & $F_{\rm S}$          & $F_{\rm S}$         & $F_{\rm S}$           \\ 
      \hline
      \rule[3.5mm]{0mm}{0mm}Stat. uncertainty                             & $^{+0.081}_{-0.073}$ & $^{+0.030}_{-0.027}$ & $^{+0.014}_{-0.007}$ & $^{+0.012}_{-0.009}$ & $^{+0.027}_{-0.025}$ & $^{+0.043}_{-0.042}$ \\[0.1ex]
      \hline
      Background subtraction  & 0.014                & 0.003                & 0.001                & 0.002                & 0.004                & 0.006                \\
      \BdToJPsiKst{} background                     & 0.010                & 0.006                & 0.001                & 0.001                & 0.002                & 0.018                \\
      Angular acc.  reweighting                     & 0.004                & 0.006                & 0.004                & 0.005                & 0.006                & 0.007                \\
      Angular acc. statistical                      & 0.003                & 0.003                & 0.002                & 0.001                & 0.003                & 0.004                \\
      Fit bias                                      & 0.009                & --                    & 0.002                & 0.002                & 0.001                & 0.001                \\
      \hline
      Quadratic sum of syst.                            & 0.020                & 0.009                & 0.005                & 0.006                & 0.008                & 0.021                \\
      Total uncertainties                            & $^{+0.083}_{-0.076}$ & $^{+0.031}_{-0.029}$ & $^{+0.015}_{-0.009}$ & $^{+0.013}_{-0.011}$ & $^{+0.028}_{-0.026}$ & $^{+0.048}_{-0.047}$ \\[0.7ex]
    \end{tabular}
  \end{center}
  \label{tab:finalsystematicssummary_FS}
\end{table}

\begin{table}[tb]
  \caption{\small Statistical and systematic uncertainties for S-wave phases in bins of $m(K^{+}K^{-})$.}
  \begin{center}\footnotesize
    \begin{tabular}{l|c|c|c|c|c|c}
      Source                                        &  bin 1   &  bin 2   & bin 3   &  bin 4   &  bin 5   &  bin 6   \\ 
      		                                        & $\delta_{\rm S}-\delta_{\perp}$    & $\delta_{\rm S}-\delta_{\perp}$  & $\delta_{\rm S}-\delta_{\perp}$  & $\delta_{\rm S}-\delta_{\perp}$  & $\delta_{\rm S}-\delta_{\perp}$   & $\delta_{\rm S}-\delta_{\perp}$   \\ 
                                                    & [rad]              &  [rad]             & [rad]              & [rad]              & [rad]              & [rad]              \\[0.5ex]
      \hline
      \rule[3.5mm]{0mm}{0mm}Stat. uncertainty                             & $^{+0.78}_{-0.49}$ & $^{+0.38}_{-0.23}$ & $^{+1.40}_{-0.30}$ & $^{+0.21}_{-0.35}$ & $^{+0.18}_{-0.26}$ & $^{+0.18}_{-0.22}$ \\[0.1ex]
      \hline
      Background subtraction  & 0.03               & 0.02               & --                  & 0.03               & 0.01               & 0.01               \\
      \BdToJPsiKst{} background                     & 0.08               & 0.04               & 0.08               & 0.01               & 0.01               & 0.05               \\
      Angular acc.  reweighting                     & 0.02               & 0.03               & 0.12               & 0.13               & 0.03               & 0.01               \\
      Angular acc. statistical                      & 0.033              & 0.023              & 0.067              & 0.036              & 0.019              & 0.015              \\
      Fit bias                                      & 0.005              & 0.043              & 0.112              & 0.049              & 0.022              & 0.016              \\
      $C_{SP}$ factors                              & 0.007              & 0.028              & 0.049              & 0.025              & 0.021              & 0.020              \\
      \hline
      Quadratic sum of syst.                            & 0.09               & 0.08               & 0.20               & 0.15               & 0.05               & 0.06               \\
      Total uncertainties                            & $^{+0.79}_{-0.50}$ & $^{+0.39}_{-0.24}$ & $^{+1.41}_{-0.36}$ & $^{+0.26}_{-0.38}$ & $^{+0.19}_{-0.26}$ & $^{+0.19}_{-0.23}$ \\[0.7ex]
    \end{tabular}
  \end{center}
  \label{tab:finalsystematicssummary_deltaS}
\end{table}

\clearpage

\section{\boldmath Results for $\BtoJpsipipi$ decays}
\label{sec:resultspipi}
The $\BtoJpsipipi$ analysis used in this paper is unchanged with respect to Ref.~\cite{LHCb-PAPER-2012-006}
except for: 
\begin{enumerate}
\item the inclusion of the same-side kaon tagger in the same manner as has already been described for the $\BtoJpsiKK$ sample. 
This increases the number of tagged signal candidates  to  2146 OS-only, 497 SSK-only and 293 overlapped events compared to 2445 in Ref.~\cite{LHCb-PAPER-2012-006}.
The overall tagging efficiency is $(39.5\pm0.7)\%$ and the tagging power increases from $(2.43\pm0.08\pm0.26)\%$ to $(3.37\pm0.12\pm0.27)\%$;
\item an updated decay time acceptance model. For this, the decay channel \mbox{$\Bd\to\Jpsi\Kstar(892)^{0}$}, which has a well known lifetime, is used to calibrate  the decay time acceptance, and simulated events are used to determine a small relative correction between the 
acceptances for the $\Bd \to J/\psi \Kstar(892)^{0}$ and $\BtoJpsipipi$ decays;
\item use of the updated values of $\Gs$ and $\DGs$ from the $\BtoJpsiKK$ analysis presented in this paper as constraints in the fit for $\phis$.
\end{enumerate}  

The measurement of $\phis$ using only the $\BtoJpsipipi$ events is 
\begin{displaymath}
\phi_s = -0.14\,^{+0.17}_{-0.16}\pm0.01\rad,
\end{displaymath}
where the systematic uncertainty is obtained in the same way as described in Ref.~\cite{LHCb-PAPER-2012-006}. 
The decay time resolution in this channel is approximately $40$\,fs and its effect is included in the systematic uncertainty.

In addition, the effective lifetime $\tau^{\rm eff}_{\Bs\to\jpsi \pi^{+}\pi^{-}}$ is measured by fitting a single exponential function
to the \Bs\ decay time distribution with no external constraints on $\Gs$ and $\DGs$ applied.
The result is
\begin{displaymath}
\tau^{\rm eff}_{\Bs\to\jpsi \pi^{+}\pi^{-}} = 1.652 \pm 0.024\ (\mathrm{stat}) \pm 0.024\ (\mathrm{syst}) \ps.
\end{displaymath}
This is equivalent to a decay width of
\begin{displaymath}
\Gamma^{\rm eff}_{\Bs\to\jpsi \pi^{+}\pi^{-}} = 0.605 \pm 0.009\ (\mathrm{stat}) \pm 0.009\ (\mathrm{syst}) \invps,
\end{displaymath}
which, in the limit $\phis=0$ and $\maglambda=1$, corresponds to $\GH$.
This result supersedes that reported in Ref.~\cite{LHCb-PAPER-2012-017}. 
The uncertainty on the \Bd\ lifetime~\cite{pdg} used to calibrate the decay time acceptance is included in the statistical uncertainty.
The remaining systematic uncertainty is evaluated by changing the background model and assigning 
half of the relative change between the fit results with and without the decay time acceptance correction included, leading to
uncertainties of $0.011\ps$ and $0.021\ps$, respectively. 
The total systematic uncertainty obtained by adding the two contributions in quadrature is $0.024\ps$. 

\section{\boldmath Combined results for  $\BtoJpsiKK$ and \mbox{$\BtoJpsipipi$} datasets}
\label{sec:resultscombined}

This section presents the results from a simultaneous fit to both  $\BtoJpsiKK$ and $\BtoJpsipipi$ datasets.  
The joint log-likelihood is minimised with the common parameters being $\Gs$, $\DGs$, $\phis$, $\maglambda$, 
$\dms$ and the tagging calibration parameters. 
The combined results are given in Table~\ref{tab-app-combined-results-II}. 
The correlation matrix for the principal parameters is given in Table~\ref{tab:corr-combined}.

For all parameters, except $\Gs$ and $\DGs$, the same systematic uncertainties as presented for the stand-alone $\BtoJpsiKK$ analysis are assigned.
For $\Gs$ and $\DGs$ additional systematic uncertainties of $0.001 \invps$  and $0.006 \invps$ respectively are included, 
due to the $\BtoJpsipipi$ background model and decay time acceptance variations  described above.

\begin{table}[tb]
  \caption{\small Results of combined fit to the $\BtoJpsiKK$ and $\BtoJpsipipi$ datasets. 
  The first uncertainty is statistical and the second is systematic.}  \begin{center}\small
    \begin{tabular}{l|c}
      Parameter            &  Value\\
      \hline
      $\Gs$ [\invps]   	& $0.661		\pm 0.004 			\pm  0.006$\\ 
      $\DGs$ [\invps]  	& $0.106		\pm 0.011			\pm 0.007$\\
      $\aperpsq$        	& $0.246 		\pm 0.007 			\pm  0.006$\\
      $\azerosq$           & $0.523 		\pm 0.005 			\pm 0.010$\\
      $\delpar$ [rad] 	& $3.32 	\,	^{+0.13}_{-0.21} \pm  0.08$\\
      $\delperp$ [rad]   & $3.04 		\pm 0.20 			\pm 0.08$\\ 
      $\phis$ [rad]         & $0.01 		\pm 0.07 			\pm 0.01 $\\
     $\maglambda$	& $0.93		\pm0.03			\pm0.02 $\\
    \end{tabular}
  \end{center}

    \label{tab-app-combined-results-II}
\end{table}

\begin{table}[t]
\caption{\small Correlation matrix  for statistical uncertainties on combined results. \label{tab:corr-combined}}
\begin{center}\small
\begin{tabular}{l|c|c|c|c|c|c|c|c} 

      					& $\Gs$ 		& $\DGs$ 		& $\aperpsq$ 	& $\azerosq$   	&$\delpar$ 	&$\delperp$ 	& $\phis$ 		&$\maglambda$	 \\ 
					                & [\invps] &  [\invps]   &    &    &  [rad]    &  [rad]  &  [rad]     &    \\\hline

                   $\Gs$ [\invps]  		&  $\phantom{+}1.00$ 		& $\phantom{+}0.10$ 	 	&  $\phantom{+}0.08$ 	 	& $\phantom{+}0.03$ 		&  $-0.08$		& $-0.04$		& $\phantom{+}0.01$			& $\phantom{+}0.00$				\\
                   $\DGs$ [\invps]  		&       		&  $\phantom{+}1.00$ 		& ${-0.49}$ 	&  $\phantom{+}{ 0.47}$ 	&  $\phantom{+}0.00$		&$\phantom{+}0.00$		&$\phantom{+}0.00$  			& $-0.01$			\\
                   $\aperpsq$ 	&       		&       		&  $\phantom{+}1.00$ 		& ${ -0.40}$  	&  ${ -0.37}$	&$-0.14$		& $\phantom{+}0.02$ 			&$-0.05$			\\
                   $\azerosq$ 	&       		&       		&       		&  $\phantom{+}1.00$ 		& $-0.05$		&$-0.03$		& $-0.01$ 			&$\phantom{+}0.01$			\\
      		$\delpar$ [rad] 		&			&			&       		&       		&   $\phantom{+}1.00$    		&  $\phantom{+}{ 0.39}$	& $-0.01$	 		&$\phantom{+}0.13$			\\
      		$\delperp$ [rad]     	&			&			&       		&       		&       		&   $\phantom{+}1.00$		& $\phantom{+}0.21$	 		&$\phantom{+}0.03$			\\
                    $\phis$	[rad]	&			&			&       		&       		&       		&   			& $\phantom{+}1.00$ 			&$\phantom{+}0.06$			\\
                    $\maglambda$&			&			&       		&       		&       		&   			& 	 			&$\phantom{+}1.00$			\\
\end{tabular}
\end{center}
\end{table}

%
%

\section{Conclusion \label{sec:conclusion}}
A sample of $pp$ collisions at $\sqrt{s}=7 \tev$,
corresponding to an integrated luminosity of $1.0$\invfb, 
collected with the LHCb detector
is used to select $27\,617 \pm 115$ \mbox{$\Bs\to\Jpsi K^{+}K^{-}$} events in a $\pm30\mevcc$
window around the $\phi(1020)$ meson mass~\cite{pdg}.
The effective tagging
efficiency from the opposite-side (same-side kaon) tagger is
${\effeff=2.29\pm0.22}$\% ($0.89\pm0.18$\%). A combination
of data and simulation based techniques are used to correct for detector efficiencies. 
These data have
been analysed in six bins of $m(\Kp\Km)$, allowing the resolution of two symmetric solutions, 
leading to the single most precise measurements of \phis, $\Gamma_{s}$ and $\DGs$
    \[
  \setlength{\arraycolsep}{1mm}
  \begin{array}{ccllllllll}
    \phi_s &\;=\; & 0.07  &\pm & 0.09  & \text{(stat)} &\pm & 0.01 & \text{(syst)} & \text{rad},\\
    \Gamma_s &\;=\; & 0.663  &\pm & 0.005 & \text{(stat)} &\pm & 0.006 & \text{(syst)} & \invps,\rule{0pt}{5mm} \\
    \DGs    &\;=\; & 0.100   &\pm & 0.016    & \text{(stat)} &\pm & 0.003 & \text{(syst)} & \invps.\rule{0pt}{5mm} \\
  \end{array}
  \]
The $\Bs\to\Jpsi K^{+}K^{-}$ events also allow an independent determination of ${\dms=17.70\pm0.10\pm0.01\invps}$.

The time-dependent   \CP-asymmetry measurement using $B_s^{0} \rightarrow J/\psi \pi^{+}\pi^{-}$ events from Ref.~\cite{LHCb-PAPER-2012-006} is updated to include same-side kaon tagger information. The result of performing a combined fit using
both $\Bs\to\Jpsi K^{+}K^{-}$ and $B_s^{0} \rightarrow J/\psi \pi^{+}\pi^{-}$ events gives
      \[
  \setlength{\arraycolsep}{1mm}
  \begin{array}{ccllllllll}
    \phi_s &\;=\; & 0.01  &\pm & 0.07  & \text{(stat)} &\pm & 0.01 & \text{(syst)} & \text{rad},\\
    \Gamma_s   &\;=\; & 0.661  &\pm & 0.004 & \text{(stat)} &\pm & 0.006 & \text{(syst)} & \invps,\rule{0pt}{5mm} \\
    \DGs  &\;=\; & 0.106   &\pm & 0.011    & \text{(stat)} &\pm & 0.007 & \text{(syst)} & \invps.\rule{0pt}{5mm} \\
  \end{array}
  \]
The measurements of $\phis$, $\DGs$ and $\Gs$ are the most precise to date and 
are in agreement with SM predictions~\cite{CKMfitter,Lenz:2006hd,*Badin:2007bv,*Lenz:2011ti}.
All measurements using  $\Bs\to\Jpsi K^{+}K^{-}$ decays supersede our previous measurements reported in Ref.~\cite{LHCb:2011aa}, 
and all measurements using $\Bs\to\Jpsi \pi^{+}\pi^{-}$ decays supersede our previous measurements reported in Ref.~\cite{LHCb-PAPER-2012-006}. 
The $\Bs\to\Jpsi \pi^{+}\pi^{-}$ effective lifetime measurement supersedes that reported in Ref.~\cite{LHCb-PAPER-2012-017}. 
The combined results reported in Ref.~ \cite{LHCb-PAPER-2012-006} are superseded 
by those reported here. Since the combined results for $\Gamma_s$ and $\Delta\Gamma_s$
include all lifetime information from both channels they should not be used in conjunction with 
the ${\Bs\to\Jpsi \pi^{+}\pi^{-}}$ effective lifetime measurement. 

\section*{Acknowledgements}

\noindent We express our gratitude to our colleagues in the CERN
accelerator departments for the excellent performance of the LHC. We
thank the technical and administrative staff at the LHCb
institutes. We acknowledge support from CERN and from the national
agencies: CAPES, CNPq, FAPERJ and FINEP (Brazil); NSFC (China);
CNRS/IN2P3 and Region Auvergne (France); BMBF, DFG, HGF and MPG
(Germany); SFI (Ireland); INFN (Italy); FOM and NWO (The Netherlands);
SCSR (Poland); ANCS/IFA (Romania); MinES, Rosatom, RFBR and NRC
``Kurchatov Institute'' (Russia); MinECo, XuntaGal and GENCAT (Spain);
SNSF and SER (Switzerland); NAS Ukraine (Ukraine); STFC (United
Kingdom); NSF (USA). We also acknowledge the support received from the
ERC under FP7. The Tier1 computing centres are supported by IN2P3
(France), KIT and BMBF (Germany), INFN (Italy), NWO and SURF (The
Netherlands), PIC (Spain), GridPP (United Kingdom). We are thankful
for the computing resources put at our disposal by Yandex LLC
(Russia), as well as to the communities behind the multiple open
source software packages that we depend on.
\appendix
\section{Definition of helicity decay angles\label{app:angles}}

The helicity angles can be defined in terms of the momenta of the decay particles.
The momentum of particle $a$ in the centre-of-mass system of $S$ is denoted by $\vec{p}_a^{\;S}$. 
With this convention, unit vectors are defined along the helicity axis
in the three centre-of-mass systems and the two unit normal vectors of the $\Kp\Km$ and $\mup\mun$ decay planes as
\begin{equation}
  \begin{gathered}
    \hat{e}_z^{\,KK\mu\mu} = +\frac{ \vec{p}_{\mup}^{\;KK\mu\mu} + \vec{p}_{\mun}^{\;KK\mu\mu} }
                                   {|\vec{p}_{\mup}^{\;KK\mu\mu} + \vec{p}_{\mun}^{\;KK\mu\mu}|},     \qquad
    \hat{e}_z^{\,KK}       = -\frac{ \vec{p}_{\mup}^{\;KK} + \vec{p}_{\mun}^{\;KK} }
                                   {|\vec{p}_{\mup}^{\;KK} + \vec{p}_{\mun}^{\;KK}|},                 \qquad
    \hat{e}_z^{\,\mu\mu}   = -\frac{ \vec{p}_{\Kp}^{\;\mu\mu} + \vec{p}_{\Km}^{\;\mu\mu} }
                                   {|\vec{p}_{\Kp}^{\;\mu\mu} + \vec{p}_{\Km}^{\;\mu\mu}|},             \\
    \hat{n}_{KK}           = \frac{\vec{p}_{\Kp}^{\;KK\mu\mu} \times \vec{p}_{\Km}^{\;KK\mu\mu}}
                                  {|\vec{p}_{\Kp}^{\;KK\mu\mu} \times \vec{p}_{\Km}^{\;KK\mu\mu}|},     \qquad\qquad
    \hat{n}_{\mu\mu}       = \frac{ \vec{p}_{\mup}^{\;KK\mu\mu} \times \vec{p}_{\mun}^{\;KK\mu\mu} }
                                  {|\vec{p}_{\mup}^{\;KK\mu\mu} \times \vec{p}_{\mun}^{\;KK\mu\mu}|}.
  \end{gathered}
  \label{eq:helicityUnit}
\end{equation}
The helicity angles are defined in terms of these vectors as
\begin{equation}
  \begin{aligned}
    \cos\thetaK  &= \frac{\vec{p}_{\Kp}^{\;KK}}{|\vec{p}_{\Kp}^{\;KK}|} \cdot \hat{e}_z^{\,KK},                  &\qquad\quad
    \cos\thetamu &= \frac{\vec{p}_{\mup}^{\;\mu\mu}}{|\vec{p}_{\mup}^{\;\mu\mu}|} \cdot \hat{e}_z^{\,\mu\mu},  \\
    \cos\phihel  &= \hat{n}_{KK} \cdot \hat{n}_{\mu\mu},                                                         &
    \sin\phihel  &= \left( \hat{n}_{KK} \times \hat{n}_{\mu\mu} \right) \cdot \hat{e}_z^{\,KK\mu\mu}.
  \end{aligned}
  \label{eq:helicity}
\end{equation}





\addcontentsline{toc}{section}{References}
\bibliographystyle{LHCb}
\bibliography{main}

\end{document}